 \documentclass[%
 amsmath,amssymb, aps, prb, twocolumn
]{revtex4}

\usepackage{graphicx,subfigure,amssymb,fancyhdr,fancyref}
\usepackage{dcolumn}%
\usepackage[toc,page]{appendix}
\usepackage{bm}%
\usepackage{color}
\definecolor{darkgreen}{rgb}{0,0.6,0}
\definecolor{darkblue}{rgb}{0,0,0.6}
\definecolor{darkred}{rgb}{0.6,0,0}
\definecolor{darkpurple}{rgb}{0.5,0,0.5}
\usepackage[colorlinks=true,urlcolor=darkblue,citecolor=darkblue,linkcolor=darkred,hyperfootnotes=false]{hyperref}

\begin{document}

\title{Distribution of zeros in the rough geometry of fluctuating interfaces}%

\author{Arturo L. Zamorategui}
 \affiliation{Laboratoire Probabilit\'es et Mod\`eles Al\'eatoires (UMR CNRS 7599), Universit\'e Pierre et Marie Curie and Universit\'e Paris Diderot, 75013 
Paris, France}
\author{Vivien Lecomte}
\affiliation{Laboratoire Probabilit\'es et Mod\`eles Al\'eatoires (UMR CNRS 7599), Universit\'e Pierre et Marie Curie and Universit\'e Paris Diderot, 75013 
Paris, France}
\author{Alejandro B. Kolton}
\affiliation{CONICET-Centro At\'omico Bariloche and Instituto Balseiro (UNCu), 8400 S.C. de Bariloche, Argentina%
}%

\date{\today}

\begin{abstract}
We study numerically the correlations and the distribution of intervals between successive zeros in the fluctuating geometry of stochastic interfaces, 
described 
by the Edwards-Wilkinson equation. For equilibrium states we find that the distribution of interval lengths satisfies a truncated Sparre-Andersen theorem. 
We show that boundary-dependent finite-size effects induce non-trivial correlations, implying that the independent interval property 
is not exactly satisfied in finite systems. For out-of-equilibrium non-stationary states we derive the scaling law describing the temporal evolution of the 
density of zeros starting from an uncorrelated initial condition. As a by-product we derive a general criterion of the Von Neumann's type to 
understand how discretization affects the stability of the numerical integration of stochastic interfaces. We consider both diffusive and  spatially fractional 
dynamics. Our results provide an alternative experimental method for extracting universal information of fluctuating interfaces such as 
domain walls in thin ferromagnets or ferroelectrics, based exclusively on the detection of crossing points.
\end{abstract}

\maketitle


\section{\label{sec:intro}Introduction}
Persistence and its related first-passage properties have been of great interest in recent years both in mathematics and in physics, going from simple models 
such as the $d$-dimensional random walk in its discrete and continuous versions\cite{redner_guide_2001, montroll_random_1965} and the non-Markovian 
acceleration process \cite{mckean_winding_1962, sinai_distribution_1992, sinai_statistics_1992}, or the $d$-dimensional Ising and Potts model at zero 
temperature with Glauber dynamics \cite{derrida_non-trivial_1994,bray_theory_1994} to  systems with many degrees of freedom such as the 
diffusion equation  of fields \cite{hilhorst_persistence_2000}, solid-on-solid surface growth models  \cite{dougherty_experimental_2002} and fluctuating 
interfaces. \cite{krug_origins_1997,  kallabis_persistence_1999, godreche_solids_1991} (For a recent 
review on persistence in non-equilibrium systems see Ref. [\onlinecite{bray_persistence_2013}].)

The notion of \textit{spatial persistence} (resp. \textit{temporal persistence}) in these systems refers to a property that does not change up to a distance 
$x$ 
(resp. up to a time $t$), being of special interest the probability distributions $P(\ell)$ and $P(\tau)$ of the corresponding distances or time intervals 
between the successive changes. Due to its possible applications, both spatial persistence 
\cite{majumdar_spatial_2001, constantin_spatial_2004, majumdar_spatial_2006} and temporal persistence~\cite{sire_crossing_2008} 
of rough fluctuating interfaces have been investigated. In this case, the property 
whose change is monitored is the sign of the interface height with respect 
to some reference line, and thus $\ell$ refers to the distances between successive zeros of an instantaneous configuration, 
and $\tau$ to the duration between successive zeros of the height at a fixed point in space. 
In a slightly different context \cite{alemany_inter-particle_1995, 
derrida_persistent_1996}, the persistent property refers to the fraction of (for instance Ising) spins that have never flipped 
up to time $t$. In order to reach a general understanding of these kind of phenomena it is useful to analyze model systems, such as 
Markovian \cite{majumdar_nontrivial_1996, derrida_persistent_1996, majumdar_spatial_2001} or non-Markovian 
\cite{constantin_persistence_2004,  sire_crossing_2008} Gaussian process with zero mean and unit variance $Y(x)$ (resp. $Y(t)$). 
In these cases the strategy to get 
$P(\ell)$ (resp. $P(\tau)$) is to extract it from the two-point correlation function $C(x,x')=\langle Y(x)Y(x') 
\rangle$ (resp. $C(t,t')=\langle Y(t)Y(t') \rangle$)  by means of 
the probability that the process $Y(x)$ (resp. $Y(t)$) has the same sign for two different points in space $x$ and $x'$ (resp. for two different times $t$ and 
$t'$ ) \cite{majumdar_nontrivial_1996, derrida_persistent_1996, majumdar_spatial_2001, sire_crossing_2008}, 
via approximation methods such as the \textit{independent interval approximation} (IIA).\cite{mcfadden_axis-crossing_1956, mcfadden_axis-crossing_1958} Even 
for some simple processes, the derivation of $C(x,x')$ (resp. $C(t,t')$) is not a trivial task. In the case of 
the Brownian motion and of the acceleration process, a suitable transformation maps these two processes into stationary Gaussian processes for which the 
two-time correlation function takes a simpler form --~it depends only on the difference between the two times considered. \cite{lamperti_semi-stable_1962} Some 
generalizations have been investigated such as the probability of having $N$ zeros between two times $t$ and $t'$ for Markovian 
\cite{majumdar_nontrivial_1996} and non-Markovian processes. \cite{majumdar_survival_1996, oerding_non-markovian_1997, sire_crossing_2008}

The persistence probability in different contexts in the large time or large distance limit is found to follow a power law  $Q(\ell)\sim \ell^{-\theta}$ 
where $\theta$ is  the persistence exponent (here again the interval $\ell$ might refer to time or spatial intervals). Even for simple diffusion it was shown 
that the persistence probability  has a nontrivial persistence exponent $\theta$. \cite{derrida_persistent_1996, majumdar_nontrivial_1996} Concerning 
fluctuating interfaces and surface growth, their persistence and their associated first-passage properties have been of considerable interest in 
the physical literature. \cite{krug_origins_1997} Starting from a flat interface and letting it evolve according to a linear Langevin fractional differential 
equation $\partial_t 
u(x,t)=-c(-\nabla ^ 2)^{\alpha} u(x,t) + \eta(x,t)$ where $\alpha$ relates to the roughness exponent $\zeta$ as $\alpha=\tfrac{1}{2}+\zeta$ (which can be 
associated with non-local harmonic elastic forces on $u(x,t)$ in general), 
it is found that the 
behavior of the temporal persistence probability $Q(t_0,t)$, understood as the probability that the interface stays above (or below) its initial value at $t_0$ 
on  the interval $[t_0,t_0+t]$,  depends strongly on the initial conditions. Two limiting cases were considered for the temporal persistence in 
Ref. [\onlinecite{krug_persistence_1997}]: for $t_0=0$, the so-called transient or coarsening persistence probability $Q_0(t)=Q(t_0=0,t)$ is found to decay as 
$Q_0(t)\sim 
t^{-\theta_0}$ for $t\to\infty$ with $\theta_0$ a non trivial exponent. On the other hand, for $t_0\to\infty$ the steady-state persistence probability 
behaves as $Q_s(t)=\lim_{t_0\to\infty}q(t_0,t)\sim t^{-\theta_s}$ with  $\theta_s=1-\zeta=\tfrac{3}{2}-\alpha\neq \theta_0$ as shown in 
Ref.~[\onlinecite{majumdar_spatial_2001}]. Extensive numerical simulations for the calculation 
of temporal persistence in surface growth processes belonging to different universality classes have been performed in 
Ref.~[\onlinecite{constantin_persistence_2004}]. The 
authors calculate separately the positive and negative persistent exponents, i.e.~the exponents associated to the intervals where the surface remains 
 above or below certain level, respectively, both transient $\theta_0^{\pm}$ and steady-state exponents $\theta_s^{\pm}$. Moreover, the authors show that 
$\theta_s^{+}$ and $\theta_s^{-}$ are always different for surfaces simulated from nonlinear equations with broken mirror symmetry, since in this case the 
surface tends to spend more time on the positive or negative values. Results have been obtained regarding spatial persistence of surface growth processes  
and for one-dimensional fluctuating interfaces.\cite{constantin_spatial_2004, majumdar_spatial_2006}

In this paper we investigate finite-time and finite-size effects for different observables 
of the stochastic dynamics of $\zeta=1/2$ Edwards-Wilkinson interfaces with periodic boundary conditions,
which can affect the statistics of its crossing zeros (we leave 
the study of the general non-Markovian case for a forthcoming paper~\cite{futureworkonnonmarkovian}).
Since accessible systems are finite, both experimentally and numerically, this kind of study 
is of importance for the numerical validation of analytical 
results and their approximations~\cite{majumdar_spatial_2006, chakraborty_finite-size_2007}.
We show in particular that finite-size effects and boundary conditions can affect
the shape of the steady-state spatial distribution of intervals $P(\ell)$, and the validity of the 
IIA for large~$\ell$. Since the Edwards-Wilkinson interface is 
linear and statistically invariant by a change of sign $u(x,t)\mapsto -u(x,t)$, 
we expect that the steady-state persistence exponents for the positive and the negative intervals are 
equal, i.e.~$\theta_s^{+}=\theta_s^{-}=\theta$. 
Further, we relate the distribution of intervals to the  first-passage distribution of a random walk. 
This mapping  between Gaussian interfaces with height $u(x,t)$ at point $x$ and time $t$ and the 
stochastic process evolving via $d^nX/dt^n=\eta(t)$ with the correspondence $u\mapsto X$ and $x\mapsto t$ is well known (see for instance 
Refs.~[\onlinecite{majumdar_spatial_2001, majumdar_airy_2005, schehr_universal_2006, majumdar_spatial_2006}]). A link between the discretized stationary 
interface and a discrete random walk is thus made by means of the Sparre-Andersen theorem. \cite{andersen_fluctuations_1953} This theorem describes the 
persistence probability $P_0(n)$ of a random walker to stay positive (or negative) up to a step 
$n$ starting in $0$. We discuss as well the influence of the boundary conditions on the correlator of 
consecutive jumps in the interface.

We also analyze the statistics of crossing points in non-stationary states, starting 
from an uncorrelated configuration. While steady states can be directly sampled 
with their equilibrium Boltzmann weight, non-stationary states are obtained by numerically  
solving the dynamics. To this end we first derive a numerically stable scheme by generalizing 
the Von Neumann stability criterion~\cite{press_numerical_2007} for 
deterministic differential equations to the general case of the 
Langevin spatially fractional differential equation describing 
interfaces with local and non-local elasticities. 
Although we present numerical results for interfaces with roughness exponent $\zeta=1/2$,
the stability condition we derive is general and relates the time step 
used for the simulations with the roughness exponent $\zeta$ and a parameter related to the time 
discretization scheme. In this context, It\=o and Stratonovich discretizations 
are just two special cases. \cite{ito_109._1944,stratonovich_topics_1967}
By solving the non-stationary dynamics within this scheme 
we obtain the scaling law describing the temporal evolution of the density 
of zeros towards the steady-state results previously analyzed.

The paper is organized as follows. In Section \ref{sec:model} we present our model and the main observables considered. 
In Section \ref{sec:stationary} we focus on the stationary state of discrete 
interfaces with periodic boundary conditions, analyzing in detail the 
finite-size effects in the distribution and spatial correlations 
of intervals between consecutive zeros.
In Section \ref{sec:nonstationary} we focus on non-stationary interfaces starting from a flat interface. 
We first compute exactly the structure factor as a function 
of time and we obtain that it keeps track of the choice of convention for the time discretization. 
From this expression  we extract a general Von-Neumann like stability criterion for 
the stochastic dynamics of interfaces. The numerical results for the evolution 
of the  structure factor of the interface and for the density of zeros yield time-dependent
scaling laws describing the approach to the steady state results.
Finally, in Section \ref{sec:conclusions} we give our conclusions and 
perspectives.

\section{\label{sec:model}Model and Observables}
In this section we introduce the model of the interface we are going to study. Although we focus in this paper on the discrete version of the system which is 
the one numerically accessible, we present first the continuous solution of the Langevin equation introduced above, for completeness. We also 
present the observables of interest, namely the length of the intervals between successive zeros, the correlation function for the intervals, the structure 
factor and the density of zeros.

To start with, we consider a fluctuating interface of size $L$ with height  $u(x,t)$ at position $x$ and time $t$ 
measured with respect to the origin. The function $u(x,t)$ satisfies the linear Langevin equation
\begin{equation}
\label{EW}
 \partial_t u(x,t)=-c(-\nabla ^ 2)^{\alpha} u(x,t) + \eta(x,t)
\end{equation}
 where the exponent $\alpha$ of the Laplacian is related to the roughness exponent $\zeta$ as $\alpha=\tfrac{1}{2}+\zeta$ and the thermal noise $\eta(x,t)$
 is defined with mean $\langle \eta(x,t) \rangle=0$ and variance $\langle 
\eta(x,t)\eta(x',t') \rangle=2T\delta(x-x')\delta(t-t')$. We consider periodic boundary conditions such that $u(0,t)=u(L,t)$. 
We will be interested in the fluctuating dynamics of the interface starting from the flat initial condition $u(x,t=0)=0$.
The general solution of Eq.~\eqref{EW} can thus be written in Fourier space as
\begin{equation}
\label{solutioncont}
 u(q,t)=\int_0^t e^{-cq^{2\alpha}(t-t')}\eta_q(t')dt'
\end{equation}
where the Fourier transform is defined as $u(q,t)=\int_{0}^{L}e^{-iqx}u(x,t)dx$, and thus 
$u(q,t=0)=0$. The Fourier noise has mean value $\langle \eta_q(t) \rangle=0$ and variance 
$\langle \eta_q(t)\eta_{q'}(t') \rangle=2TL \delta(t-t')\delta_{qq'}$. We will be interested both in the 
non-stationary and the steady-state solutions of Eq.~\eqref{EW} which is reached at long times 
$t \sim L^{2\alpha}$. 

At a given time $t$, the average height is  $\bar{u}(t)=\frac{1}{L}\int_0^{L}u(x,t)dx$ where $L$ is the size of the interface. We define a zero as the crossing 
point of the interface with its mean value $\bar{u}(t)$, i.e.~the points such that $u(x,t)-\bar{u}(t)=0$. 
Although the Fourier modes are independent, as shown by (2), the zeros are defined in real space and thus  
have a non-trivial statistics. For simplicity we can fix the mean value of the 
interface to zero, which is equivalent to fixing the amplitude of the first mode to zero $u(q=0,t)=0$. Under this assumption a zero of $u(x,t)$ is identified 
with a change of sign.

\begin{figure}[h!]
\centering
\includegraphics{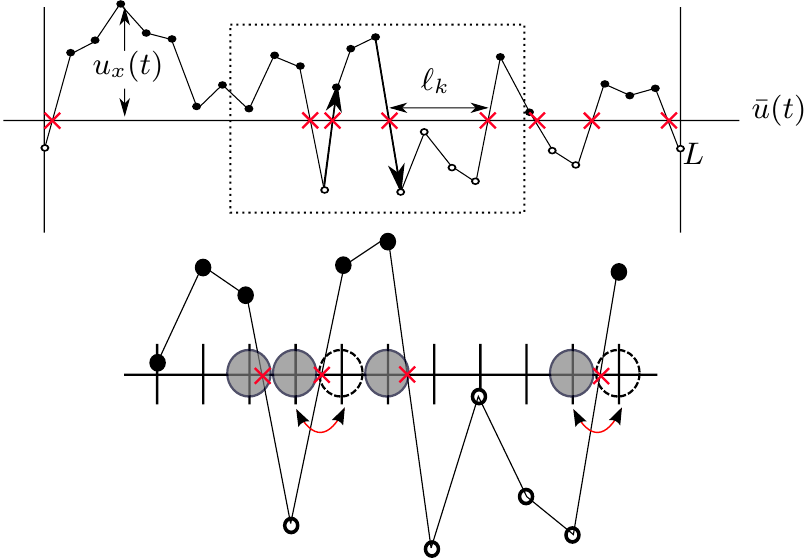}
\caption{An interface with periodic boundary conditions and height $u_x(t)$ measured with respect to its average value  $\bar{u}(t)$ defined on a lattice with 
$L$ sites.  The zeros are the points where the interface intersects its mean value  $\bar{u}(t)$ given by the crosses. Such zeros divide the lattice in $N$ 
intervals of lengths $\ell_i$, $i=1,\dots,N$.  Since we work in discrete space, the detection of the zeros can be either selecting the site on the left 
of the point where the interface crosses its average value (filled circles at the bottom) or by choosing the site that is closer to the 
crossing point (empty circles). Notice how the choice of the method modifies the length of the intervals between consecutive zeros. In this case the second and 
the fourth gray zero are on the next site on the right if the nearest site was chosen. Moreover, when choosing a zero as the nearest site to the crossing 
point, one can find two zeros at the same site which in turn allows the existence of intervals of length $\ell=0$.}
 \label{interface}
\end{figure}

In the following  we work on a lattice of $L$ sites with spacing $\Delta x=1$ and we denote the height of the interface as $u_x(t)$ with $x=1,2,\dots,L$ (See 
Fig.~\ref{interface}). We focus on the Fourier transform of the height defined as $u_{q_k}(t)=\sum_{x=0}^{L-1} u_x(t) e^{- i q_k x}$, with $q_k=2\pi k/L$ and 
$k=0,\dots,L/2+1$, as an abuse of notation we will omit the subscript $k$ in the following. $u_{q}(t)$ satisfies the general discretized equation 
\eqref{fourieruGen} introduced below. Periodic boundary conditions and $u_{q=0}(t)=\sum_{x=0}^{L-1} u_x(t)=0$ are assumed. The identification of the zeros of 
the interface on a lattice is not trivial as in the continuous case. In Section \ref{subsec:persistenceRW} we discuss in detail such way of detecting the zeros 
which in turn is fundamental for a proper description of the intervals. After detecting a change of sign of the interface from one site to the next, an 
appropriate method must be defined so that there is not ambiguity in choosing what site contains the zero (see Fig.~\ref{interface}). In this paper we define a 
zero to be the site immediately to the left of the crossing at which the height changes sign. The intervals between consecutive zeros are denoted by $\ell_i$ 
with $i=1,\dots,N$ and by $N$ the number of intervals.

In the stationary state of the interface, we look at the distribution of the lengths of the spatial intervals $P(\ell)$, which will be defined carefully in 
the next section. In the following, when talking about an interval of the interface we refer to the length $\ell$ of a spatial interval defined on the lattice 
with spacing $\Delta x=1$. The distribution of the intervals $\ell$ is obtained by direct sampling of stationary configurations. We will also study the spatial 
correlations of the intervals where the correlation function is defined as $C(r)=\langle \ell_i \ell_{i+r} \rangle - \langle \ell_i \rangle \langle  
\ell_{i+r}\rangle$ for two intervals $\ell_i$ and $\ell_{i+r}$ averaged over all the (ordered) pairs of intervals. Additionally, we discuss how periodic 
boundary conditions induce correlations in the jumps even for a random 
walk with increments $\eta_x$ with mean $\langle \eta_x \rangle=0$ and variance $\langle \eta_x\eta_x^{\prime} \rangle=2T\delta_{xx'}$. In the non-stationary 
state we are interested in the evolution of the density of zeros $\rho(t)=N/L$, where $N$ is the number of zeros and $L$ is the size of the lattice. Such 
density of zeros is extracted from the dynamics of the interface evolving from a flat interface at time $t=0$. Concerning the interface, we analyze the 
structure factor defined as $S_q(t)=\langle u_q^*(t)u_q(t)\rangle$ where the $u_{q}(t)$ is the Fourier transform of $u_x(t)$, with $q=2\pi k/L$ and 
$k=0,\dots,L/2+1$. 

\section{\label{sec:stationary} Stationary state}
To start with, we analyze the stationary features of the interface. In this section we refer to the height of the interface at a certain point $x$ 
simply as $u_x=u_x(t\to\infty)$ by omitting the time dependence. We begin with a brief discussion about the persistence properties of a 
discrete random walk which will be naturally extended to the distribution of the intervals between consecutive zeros. Later on, we study numerically the 
correlations between such intervals and we present a scaling function for such correlations. We conclude this Section by describing how the boundary conditions 
are determinant for the appearance of correlations. In particular we look at the correlation of the spatial increments of the interface.

\subsection{\label{subsec:persistenceRW} Distribution of intervals as first-passage distribution of a random walk.}
Let us denote $u_n$ the position of an unbiased random walker at step $n$. Then, the 
persistence probability for this random walker to stay positive up to step $n$, having started in $u_0>0$, is denoted by $Q(u_0, n)$. Similarly, the 
 probability that the random walker reaches the origin in exactly $n$ steps starting in $u_0>0$ is $P(u_0, n)=Q(u_0, n)-Q(u_0, n+1)$, which is known as the 
first-passage probability. When the random walk is defined in continuous space and time  $u(t)$ with $t\geq 0$, the first-passage probability 
$P(t)$ is defined as the probability density of the time at which the random walker changes sign, i.e.~the probability that the process has a zero at 
time $t$. In this case $P(t)=-dQ(t)/dt$, with $Q(t)$ the persistence probability that the walker stays positive between time $0$ and time $t$. Note that 
already for this simple process in the discrete case the definition of a zero is not trivial. We will discuss in detail how to detect the zeros when the 
process is discrete when we describe the relation of the discrete random walk with the fluctuating interface.

Consider now the jump distribution of the random walk given by the function $\phi(\eta)$ which we assume symmetric 
and continuous. The persistence probability $Q(u_0, n)$ is the probability that $u_i\geq 0$ for all $i=1,\dots,n$ having started in $u_0$. By 
considering the first step of the walker with a stochastic jump from $u_0$ to $u_1$ and letting evolve the random walker for $n-1$ steps with the 
jumps $\eta_i=u_{i}-u_{i-1}$  being independent and identically distributed, we can write a backward equation for $Q(u_0, n)$ as 
\begin{equation}
\label{persistence}
 Q(u_0,n)=\int_0^{\infty}  Q(u_1,n-1)\phi(u_1-u_{0})du_1
\end{equation}
with initial condition $Q(u_0,0)=1$ for all $u_0\geq 0$.

Although $Q(u_0,n)$ depends explicitly on the jump distribution $\phi(\eta)$, as seen in Eq.~\eqref{persistence}, 
it can be shown that under our previous assumptions $Q(0,n)$ is independent of  $\phi(\eta)$ 
(See Ref.~[\onlinecite{sire_crossing_2008}] and references therein). 
Moreover $Q(0,n)$ takes the following simple 
form
\begin{equation}
\label{sparreandersen}
Q(0,n)=\binom{2n}{n}2^{-2n}
\end{equation}
which is the celebrated Sparre-Andersen theorem. \cite{andersen_fluctuations_1953}

In the limit of large $n$, the persistence probability $Q(0,n)$ behaves as $\sim n^{-\theta}$ with $\theta=1/2$ the persistence exponent. This exponent is 
universal in the sense that even when $Q(u_0,n)$ depends on $\phi$, $\theta$ does not depend neither on $u_0$ nor on $\phi$. It is easy to see that the  
first-passage probability $P(0,n)$ behaves as $\sim n^{-3/2}$ at large~$n$.

The excursion made by a discrete random walker starting at $u_0=u_{n=0}$  that remains positive up to step $n$ and becomes negative at step $n+1$ resembles
the behavior of the interface in a given interval starting from $u_0=u_{x=0}(t=0)$.  With this image in mind, we 
can think of the number of steps $n$ during which the random walker does not change sign as the length $\ell$ of the intervals generated by the zeros of 
the interface (See Fig.~\ref{interface}). This suggests that the persistence probability 
$Q(u_0, n)$ defined for the random walk might describe well the probability $Q(u_0, \ell)$ that the interface $u_x$ stays above its mean height for a 
distance $x=\ell$. At this point, even if this idea seems reasonable, we are not ready to extend the first-passage probability $P(n)$ of the discrete random 
walk to the probability $P(\ell)$ that the interface has an interval of length $\ell$. First let us discuss how to define a zero when working on a 
lattice. Note that in this case a zero can no longer be identified just by a change of sign of $u_x$ (See Fig.~\ref{interface}). Let us imagine that 
$u_{x-1}(t)>0$ at $x-1$, 
$u_x<0$ on the next site and $u_{x+1}>0$ again on the next, then it exists $x^{\prime}\in\mathbb{R}$ in the interval $(x-1,x)$  such that $u_{x'}(t)=0$  and 
$x''$ in the interval $(x,x+1)$ such that $u_{x''}=0$. Moreover let us assume that both $x'$ and $x''$ are closer to $x$, in this case we could define a zero 
to be the site on the lattice that is closer to the crossing point. If this were the case, we would find two zeros on the same site $x$ in our example, 
therefore an interval of length $\ell=0$ is found between these two zeros. Another way of finding the zeros would be choosing always the site on the left (or 
on 
the right) to the crossing point, this would leave in this case a zero on the site $x-1$ and other zero on $x$, and thus an interval of length 
$\ell=1$ between these zeros. The later method, which we will adopt, produces more simple results for which the Sparre-Andersen theorem also applies in spite 
of the small correlations induced by periodicity.

Always choosing the site on the left (by symmetry we obtain the same results if we choose the 
site on the right) prevents us of having intervals of length $\ell=0$ which however are allowed in the calculation of the persistence function $Q(u_0,n)$ 
 of a random walk. \cite{krug_persistence_1997} In our case, if such intervals of length $\ell=0$ were allowed, we observe, in comparison with the 
Sparre-Andersen theorem, that the statistics for intervals of length $\ell=1$ changes and becomes more sensitive to the correlations induced by periodic 
boundary condition, that are present even at short scales (see subsection~\ref{subsec:correlations}). Another issue to take into account is that 
Eq.~\eqref{sparreandersen} is obtained by choosing the initial condition of the random walker as 
$u_0=0$. However, for our interface with continuous displacements it is rare to have an interval that starts exactly at $u_0=0$. 
The effect of the discretization for a random walk on a semi-infinite domain produces that the average 
return time to the origin (first-passage time) is finite. This is in contrast with the continuous random walk for which such mean interval is zero 
due to the infinite number of crossings that follows after the random walker changes sign for the first time before making a long excursion.  
Moreover, the periodic boundary conditions constrain the sum of the lengths of the intervals to be exactly $L$ the size of the lattice given, i.e.~$\sum_k^N 
\ell_k=L$, with $N$ the number of intervals which is always even. 
Below we discuss the influence of large intervals of lengths comparable to the system size ($\ell\sim L/2$) on 
the tail of the distribution. This question also appears in the study of extremal or record statistics in random systems as studied in 
Refs.~[\onlinecite{frachebourg_extremal_1995, 
godreche2015record}]. Analogously, periodic boundary conditions, which turn out to be crucial in the study of the steady-state distributions for finite 
interfaces, are also considered in Ref.~[\onlinecite{majumdar_airy_2005, majumdar_spatial_2006, schehr_universal_2006}].

Under the previous assumptions, the probability $p(\ell)$ of having an interval of length $\ell\geq 1$  is in a good agreement with the 
first-passage probability of a discrete random walk with initial condition $u_0=0$ after normalization which gives a modified 
Sparre-Andersen theorem 
\begin{equation}
\label{modsparreandersen}
 p(\ell)=\frac{1}{Z}(Q(\ell)-Q(\ell+1))
\end{equation} 
Here $Q(\ell)=\binom{2\ell}{\ell}2^{-2\ell}$ as in Eq.~\eqref{sparreandersen} and $Z$ is a normalization factor
\begin{align}
\nonumber
 Z=\sum_{\ell=1}^{\ell_{\text{max}}}P(\ell)&=\sum_{\ell=1}^{\ell_{\text{max}}}[Q(\ell)-Q(\ell+1)]\\
\label{normalprobano0}
 &=\frac{1}{2}-2^{-2L}\binom{2L}{L}
\end{align}
which rules out intervals of length $\ell=0$ and $\ell>\ell_{\text{max}}=L-1$.

\begin{figure}[h!]
\centering
\def\svgwidth{260pt}
\includegraphics{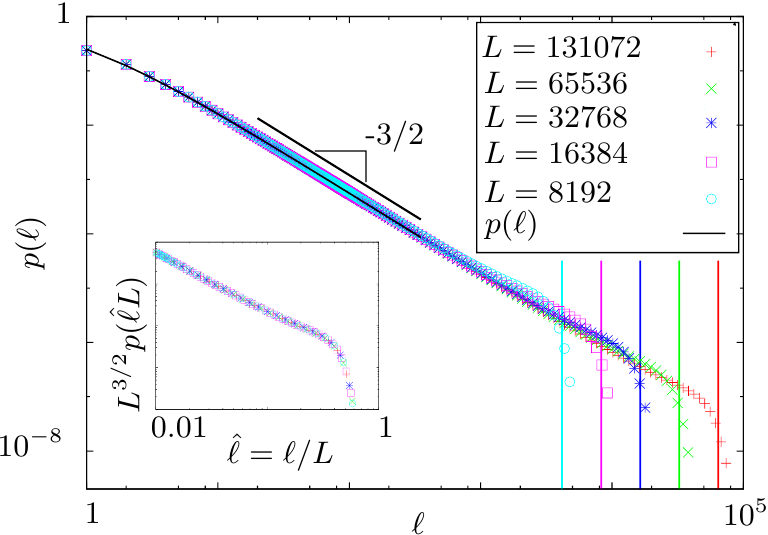}
\caption{Normalized histogram of the distances between consecutive zeros $\ell$ for different sizes of the system compared to the modified Sparre-Andersen 
theorem $p(\ell)=(q(\ell)-q(\ell + 1))/Z$  with $Z$ the normalization factor given by  Eq.~\eqref{normalprobano0} and 
$q(\ell)=\binom{2\ell}{\ell}2^{-2\ell} $ which is in good agreement even for small values of $\ell$. Vertical lines show the values of $L/2$ for $L=8192, 
16384, 32768, 
65536, 
131072$ (in cyan, fuchsia, blue, green and red respectively, in the color version). Inset shows the tail of the distribution for $L$ large (which are 
the 
points for which Sparre-Andersen is no longer valid) with its points calculated from the average of the original data taken on logarithmic bins. In the limit 
$L\to\infty$ Sparre-Andersen is always satisfied. A scaling law is found for the probability of the intervals for large values of $\ell$ for which all the 
curves superpose. This law is $p(\ell,L)\sim L^{-3/2} \hat{p}(\ell/L)$ as explained in the text . }
\label{pdfell}
 \end{figure}

By direct-sampling of stationary configurations we can obtain numerically the distribution of the intervals for different system sizes $L$ (see 
Fig.~\ref{pdfell}). The stationary configurations in Fourier space for an interface with roughness exponent $\zeta$ present a Gaussian distribution and can 
thus be directly obtained from the structure factor $S_q=S_q(t\to\infty)=\langle u^{*}_{q}u_{q} \rangle$.

For our Edwards-Wilkinson interface of interest, we have used the large-scale expression $S_q=\frac{TL}{cq^{2}}$ with $TL=\tilde T=0.1$ and $c=1$.
We discuss  in Section~\ref{sec:nonstationary} other choices of structure factors including e.g.~finite-size corrections.

This is done by generating random Gaussian amplitudes $u_q$  with zero 
mean and variance proportional to $S_q$ as explained in Ref.~[\onlinecite{krauth_statistical_2006}]. We find 
that the histograms of the intervals $\ell$ for different $L$ satisfy, up to corrections due to the discretization, the modified Sparre-Andersen 
theorem~\eqref{modsparreandersen} as shown in Fig.~\ref{pdfell}, at least in the region where the length of the intervals is much smaller than~$L/2$. Two 
comments are in order. The method of defining the zeros discussed in the previous subsection influences the results for $\ell>1$: in fact, our convention for 
the definition of the location of zeros proves to be in surprising good agreement with the Sparre-Andersen theorem, with the normalization factor described 
in~\eqref{normalprobano0}.
One could expect that correlations induced by the periodic boundary conditions would render this result invalid, but this is not the case. However, if one 
takes other definitions for the locations of zeros, the correspondence would not hold.  Second, for large $\ell$ but smaller than the system size, in 
particular 
for $\ell<L/2$, the intervals satisfy a power law behavior $P(\ell)\sim \ell^{-\gamma}$ with $\gamma = 3/2$. Since this exponent $\gamma$ for the distribution 
of intervals is related to the steady-state persistence exponent $\theta$ as $\gamma=\theta+1$, we obtain the expected persistence exponent $\theta=1/2$ for 
fluctuating interfaces with roughness exponent $\zeta=1/2$ or the persistence exponent of a discrete random walk.
 
For $\ell$ above $L/2$, the effects of periodicity are very strong and this induces a cut-off at this lengthscale as observed in the numerical results.  A 
corresponding scaling law can in fact be found for the distribution of the intervals for large values of $\ell$ which turns out to be $p(\ell)\sim L^{-3/2} 
\hat{p}(\ell/L)$, with $\hat{p}(x)$ decaying rapidly for $x \gg 1$, and $\hat{p}(x) \sim x^{-3/2}$ for intermediate $x\ll1$. 
Fig.~\ref{pdfell} shows that intervals of length $\ell>L/2$ are indeed very rare, 
it also shows that the scaling is valid in the region for large $\ell<L$. 

\subsection{\label{subsec:correlations} Correlations of increments}
For a 1d interface drawn from the steady state described with height $u_x$  at a distance $x$ with initial condition $u_0=0$, the 
consecutive increments are decorrelated since the process is Brownian along the spatial direction $x$ and thus Markovian. By comparing two simple examples, we 
investigate first how periodic boundary 
conditions induce correlations on the increments even for this simple process. We start first with the analysis of the correlator of the increments of a random 
walk attached in one end and free in the other to find afterwards the correlator of the process itself. Then we impose boundary conditions and find the 
correlator of the jumps for this constrained system which turn out to present long-range correlations.  We underline that the discretization in space and the 
periodic boundary conditions help us to understand the correlations between intervals. By the end of this Section we define the correlation function 
for the intervals in the stationary state and present some numerical results.

\subsubsection{Example: Interface attached in one end and free in the other.}
In this case the interface is attached at the origin $u_0=0$ but it is free at the other extreme. The height at every position is determined by $u_{x+1}=u_x+ 
\eta_x$ where the noise is distributed as $\langle\eta_x\rangle=0$ and second moment $\langle \eta_x \eta_{x'}\rangle=2T \delta_{xx'}$, thus the 
jumps are uncorrelated. Therefore the process at position $x$ is defined as
\begin{equation*}
 u_x=\sum_{x^{\prime}=0}^{x-1} \eta_{x^{\prime}}
\end{equation*}
or
\begin{equation}
\label{delta}
  \vec u = \mathbb{M} \vec\eta,  
\end{equation}
with $\mathbb{M}$ a lower triangular matrix.

The probability of a history is obtained from that of the noise, which is Gaussian, and for our choice of correlations above reads as follows
\begin{equation}
\label{probadelta}
 P(\eta_0,\eta_1,\dots,\eta_{L-1})\propto \exp (-\tfrac{1}{2T}\vec\eta\, \mathbb{K}_{\eta_x} \vec\eta)
\end{equation}
with $T$ the physical temperature of the process and $\mathbb{K}_{\eta_x}$ the identity matrix of dimension $L\times L$.
To see how the process $\vec u$ is distributed we can get $\vec \eta$ from Eq.~\eqref{delta} and insert it in Eq.~\eqref{probadelta} as 
follows
\begin{align*}
 P(\vec u) &\propto \exp (- \tfrac{1}{2T}\vec u (\mathbb{M}^{-1})^{\dagger}\mathbb{K}_{\eta_x} \mathbb{M}^{-1}\vec u)\\
	  & = \exp(-\tfrac{1}{2T}\vec u\, \mathbb{K}_{u_x} \vec u) \\
	  & = \exp\Big(-\tfrac{1}{2T}\sum_{x=0}^{L-1}(u_{x+1}-u_x)^2\Big)
\end{align*}
since the transformation \eqref{delta} has unit Jacobian. The matrix $\mathbb{K}_{u_x}=(\mathbb{M}^{-1})^{\dagger}\mathbb{K}_{\eta_x} \mathbb{M}^{-1} $ has the 
form of a discrete Laplacian
\begin{equation*}
  \mathbb{K}_{u_x} = 
  \begin{pmatrix}
 2 & -1 & 0 & \cdots & 0 \\
 -1 & \ddots & \ddots & \ddots & \vdots \\
  0 & \ddots & \ddots & \ddots & 0\\
 \vdots & \ddots & \ddots & 2 & -1 \\
 0 & \cdots & 0 & -1 & 1
\end{pmatrix}.
\end{equation*} 

The Fourier transform of $\vec u$ can be also expressed in matrix form as $ \vec u_q = \mathbb{F} \vec u_x$, where we leave the subindex to identify 
$\vec u$ from its Fourier transform. The elements of the matrix $\mathbb{F}$ are $\mathbb{F}_{j k}=\exp(-\tfrac{2i \pi}{L}jk)$
and the elements of its inverse are $(\mathbb{F}^{-1})_{jk}=\frac{1}{L}\exp(\tfrac{2i\pi}{L}jk)$
Similarly, we can define the Fourier transform of the noise as $ \vec\eta_q=\mathbb{F}\vec\eta_x$.

\subsubsection{Example. Interface with periodic boundary conditions}

To determine the correlations induced by the periodic boundary conditions (PBCs), we now follow the procedure of the previous example, but backwards. We will 
start from the known distribution of the interface position and deduce from it the correlator of its elementary increments~$\eta_x$. For PBCs, the 
probability of a history in Fourier space $\vec u_q$ is (See Section \ref{sec:nonstationary} for details):
\begin{align}
 \nonumber
 P[\vec u_q]&=\exp(-\frac{1}{2T}\sum_q 2(1-\cos q)u^*_q u_q)\\
 \label{un1}
  & = \exp(- \tfrac{1}{2T}\vec u^*_q \mathbb{K}_q \vec u_q)
\end{align}
with $q=\tfrac{2\pi k}{L}$, $\vec u_q=\mathbb{F}\vec u_x$ the Fourier transform of $\vec u_x$ and $\sum_q$ running over the Fourier index $k=0\ldots L-1$.

The probability of a history $\vec u_x$ with PBCs takes a similar form as in the previous example. We substitute $\vec u_q$ in Eq.~\eqref{un1} in terms of 
$\vec u_x$ as
\begin{align}
 \nonumber
 P^{\text{PBC}}[\vec u_x]& \propto \exp(-\tfrac{1}{2T}\vec u_x\mathbb{F}^{\dagger}\mathbb{K}_q\mathbb{F}\vec u_x) \\
 \label{ux}
 & = \exp(-\tfrac{1}{2T}\vec u_x \mathbb{K}_{u_x}^{\text{PBC}}\vec u_x)
\end{align}
where
\begin{equation*}
\mathbb{K}_{u_x}^{\text{PBC}}=
   \begin{pmatrix}
 2 & -1 & 0 & \cdots & 0& -1 \\
 -1 & \ddots & \ddots &\ddots &  & 0 \\
  0 & \ddots & \ddots & \ddots & \ddots & \vdots\\
  \vdots & \ddots & \ddots & \ddots & \ddots & 0\\
 0 &  & \ddots & \ddots & \ddots& -1 \\
 -1 & 0 & \cdots & 0 & -1 & 2
\end{pmatrix}.
\end{equation*}

\begin{figure}[h!]
\centering 
\includegraphics{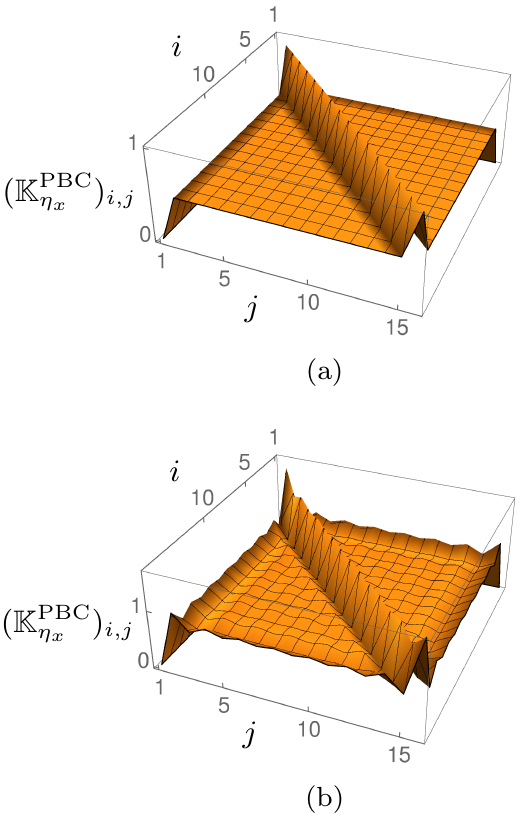}
\caption{\small (a) Correlator of the noise $\mathbb{K}^{\text{PBC}}_{\eta_x}$ of an interface with periodic boundary 
conditions using the exact Laplacian. (b)  Correlator of the noise $\mathbb{K}^{\text{PBC}}_{\eta_x}$ of an interface with periodic boundary conditions 
with the approximate Laplacianµ $(1-\cos q )\approx \tfrac{1}{2}q^2$ (see Eq.~\eqref{un1}). 
The increment correlations out of the diagonal persist even if we cut some Fourier modes.}
\label{matrixKdx}
\end{figure}

Then in Eq.~\eqref{ux} we express $\vec u_x=\mathbb{M}\vec\eta_x$ in terms of $\vec\eta_x$ as follows
 \begin{align*}
  P^{\text{PBC}}[\vec\eta_x] & \propto \exp(-\tfrac{1}{2T}\vec \eta_x(\mathbb{F}\mathbb{M})^{\dagger}\mathbb{K}_q^{\text{PBC}}\mathbb{F}\mathbb{M}\vec \eta_x)\\
	& = \exp(-\tfrac{1}{2T}\vec \eta_x\mathbb{K}_{\eta_x}^{\text{PBC}} \vec \eta_x)
 \end{align*}
where
\begin{equation*}
\mathbb{K}_{\eta_x}^{\text{PBC}}=
\begin{pmatrix}
 0 & \cdots & \cdots & \cdots  & 0 \\
 \vdots & 2 & 1 & \cdots & 1 \\
  \vdots & 1 & \ddots & \ddots  & \vdots\\
 \vdots & \vdots & \ddots & \ddots & 1 \\
 0 & 1  &\cdots & 1 & 2
\end{pmatrix}
\end{equation*}
describes the correlations of the increments of our interface with PBC.~(see Fig.~\ref{matrixKdx}(a)). We thus observe that the increments present long-range 
correlations as a result of the boundary conditions. Note that if we take the approximation of the Laplacian for small values of $q$ in 
Eq.~\eqref{un1}, i.e.~$1-\cos q\approx \frac{1}{2}q^2$, we also obtain a matrix $\tilde{\mathbb{K}_{\eta_x}^{\text{PBC}}}$ presenting long-range correlations 
for the increments, as shown in Fig.~\ref{matrixKdx}(b).

\subsection{\label{subsec:correlationNumerics}Spatial correlation of intervals}

To see how the intervals generated by the zeros are correlated we compute the correlation function
 \begin{equation}
 \label{correlation}
  C(r)=\langle \ell_i \ell_{i+r} \rangle - \langle \ell_i \rangle \langle  \ell_{i+r}\rangle,
 \end{equation}
 where the average is made over all the $N$ intervals of the interface. 

\begin{figure}[h!]
 \centering
\includegraphics{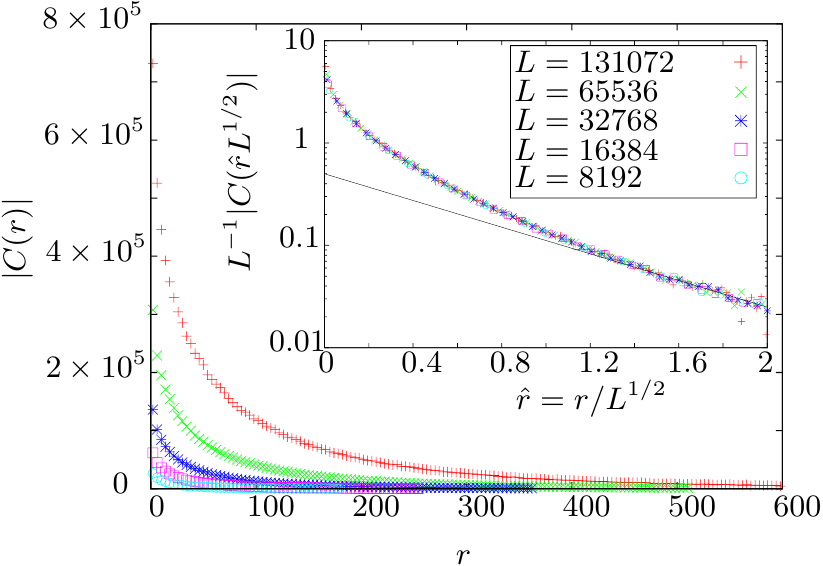}
\caption{ Absolute value of the interval correlation $C(r)$ as defined in Eq.~\eqref{correlation} for different sizes $L$ of the system and $r$ even. We show 
the rescaled correlation in log-normal scale which follows a scaling law $|C(r)|=L |\hat C(r/L^{1/2})|$. The obtained scaling function $\hat C$ is independent 
of $L$ (inset 
figure). $|C(r)|$ seems to follow a non-trivial law as compared to a pure exponential law $\sim e^{- 3\hat{r} / 2}$ given by the solid line in the 
inset.}
\label{intercor1}
 \end{figure}

For the correlation we observe that for odd values of $r$, the intervals are weakly anti-correlated for small values of $r$ but converge to zero almost 
immediately. However, for even values of $r$ the intervals are strongly anti-correlated (except for $r=0$) and tend more slowly to zero as observed in 
Fig.~\ref{intercor1} where we plot $|C(r)|$ for $r$ even. The decay of the correlations seems to approach an exponential behavior at large~$r$. Moreover the 
correlation function obeys a scaling  law $|C(r)|=L |\hat C(r/L^{1/2})|$ as illustrated in the inset of Fig.~\ref{intercor1}.

\section{\label{sec:nonstationary}Non-stationary state}
 The simplest equation to describe the evolution in time of a rough interface is the well-known Edwards-Wilkinson (EW) equation defined in continuous space and 
time in Eq.~\eqref{EW}. In this section we will focus on a general discretized version of this equation and  obtain a 
 stability criterion that generalizes the well-known Von Neumann stability \cite{press_numerical_2007}. This criterion establishes the necessary condition for 
the solution to be stable given the discretization scheme chosen.

We denote $u_q(t)$ as the Fourier transform of the height of the interface 
$u_x(t)$, here again we write explicitly the time dependence.  For the discretization of the EW equation, the time derivative can be discretized by taking a 
proportion $a\,u_q(t)$ of the function at the current step 
and $(1-a)u_q(t+\Delta t)$ of the function one step later, with $a\in [0,1]$. This results in a general form of the discretized EW equation as follows
\begin{equation}
 \label{fourieruGen}
  u_q(t+\Delta t) = \frac{[1-a K_q] u_q(t)  + \sqrt{\tilde T\Delta t}\,\eta_q(t)}{1 +(1-a) K_q }
\end{equation}
where $\tilde T=TL$ and  $K_q= c \Delta t(2(1- \cos q))^{\alpha}$ the Laplacian with roughness coefficient $\zeta=\alpha-\tfrac{1}{2}$. $K_q$ can 
be approximated as $c\Delta t q^{2\alpha}$ for small values of $q$.

In this equation we recognize the It\=o and Stratonovich discretization  when choosing $a=1$ and $a=1/2$, respectively. As we will discuss below, the choice
of the time-discretization parameter $a$ is rather important: it influences the form of steady-state itself, and the stability of the numerical scheme.

\subsection{Discrete-time solution}
The solution $u_q(t)$  of the continuous EW equation in Fourier space (See Eq.~\eqref{EW}) given by the expression \eqref{solutioncont} is obtained by direct 
integration of Eq.~\eqref{EW}. By introducing $v_q(t)=f_q(t)u_q(t)$ such solution is simply the integral of 
$\partial_t(v_q(t))=\partial_t(f_q(t)u_q(t))=e^{cq^2t}\eta_q(t)$. Notice that by considering a step $\Delta t$ we have that $v(t+\Delta t)-v(t)\approx \Delta t 
e^{cq^2t}\eta_q(t)=g(t)$  with $g(t)$ a function that does not depend explicitly on $u_q(t)$. In the following we find a similar solution for the 
discrete equation \eqref{fourieruGen} written as a geometric sum as in Equation \eqref{solutiondiscrete} below. We provide the details of the computations 
since it allows to pinpoint the precise origin of the stability from the convergence of a geometric sum.

Equation \eqref{fourieruGen} corresponds to the most general discretization of the EW equation in Fourier space for which the  discretization is controlled by 
the parameter $a$. Let us rewrite Eq.~\eqref{fourieruGen} as follows 
 \begin{equation}
 \label{fourieruGenshort}
   u_q(t+\Delta t) =A_q u_q(t)  +  B_q(t),
 \end{equation}
with $A_q$ and $ B_q(t)$ given by 
 \begin{align*}
  A_q=\frac{1-aK_q}{1 +(1-a)K_q}
 \end{align*}
 and  
 \begin{equation*}
  B_q(t)=\frac{ \sqrt{\tilde T\Delta t}\,\eta_q(t)}{1 
+(1-a) K_q}
 \end{equation*}
where $a\in[0,1]$ and $K_q= c \Delta t(2(1- \cos q))^{\alpha}$ or for small values of $q$, $K_q= c \Delta t q^{2\alpha}$  with $\alpha=\tfrac{1}{2}+\zeta$ 
for any roughness exponent $\zeta$.  

We would like to find $f_q(t)$ so that we can express $v_q(t)=f_q(t)u_q(t)$ as for the continuous case and from here to find $u_q(t)$ that satisfies Eq.
\eqref{fourieruGenshort}.

To start with, we seek a function  $g(t)$  independent of $u_q(t)$ such that $v_q(t+\Delta t)-v_q(t)=g(t)$. This holds if $A_qf(t+\Delta t)-f(t)=0$, 
i.e.~$f(t)=A_0A_q^{-t/\Delta t}$. Notice that $f(t)$ can also be written as $f(t)=A_0e^{-\tfrac{t}{\Delta t}\log \tfrac{1-aK_q}{1+(1-a)K_q}}\approx A_0e^{tc 
q^2 + \mathcal{O}(\Delta t)}$ (we used the approximation for small values of $q$ and $\zeta=1/2$). Thus, in this case we recover the function $f(t)$ 
found in the continuous solution of the EW equation by taking the limit $\Delta t\to 0$.

For finite $\Delta t$, one has
\begin{equation*}
 v_q(t+\Delta t)-v_q(t)=A_q^{-t/\Delta t}\eta_q(t)
\end{equation*}
with $A_0=\frac{1+(1-a)K_q}{\sqrt{\tilde T\Delta t}}A_q$.
Then we can find the solution of $v(t)$ as follows
\begin{equation*}
v_q(t)=\sum_{t^{\prime}=0}^{t-\Delta t} A^{-t^{\prime}/\Delta t}\eta_q(t^{\prime}) + A_0v_q(0)
\end{equation*}
with the step in the sum of size $\Delta t$.
The solution $u_q(t)=v_q(t)/f_q(t)$ in discrete time is
\begin{equation}
\label{solutiondiscrete}
u_q(t)= \frac{1}{A_0}\sum_{t^{\prime}=0}^{t-\Delta t}  A_q^{(t-t^{\prime})/\Delta t}\eta_q(t^{\prime}) +  A_q^{t^{\prime}/\Delta t}u_q(0).
\end{equation}
where we take $u_q(0)=0$ for simplicity. This is the discrete equivalent of the continuous solution \eqref{solutioncont}.

\subsection{\label{subsec:structureTime}Structure factor from discrete-time solution}
We found above the discrete solution of Eq.~\eqref{fourieruGen}. From here the structure factor can be found straightforwardly as follows
\begin{equation}
 S_q(t)=\langle u^*_q(t)u_q(t)\rangle=\frac{2}{A_0^2} \sum_{t^{\prime}=\Delta t}^{t} A_q^{2t^{\prime}/\Delta t}
 \label{eq:Sqgeomseries}
\end{equation}
where we used the fact that $\langle\eta_q(t)\eta_q(t^{\prime})\rangle=2\delta_{tt^{\prime}}$.

Hence,
\begin{equation*}
 S_q(t) = \frac{2}{A_0^2}\frac{A_q^2(1-A_q^{2t/\Delta t})}{1-A_q^2}.
\end{equation*}
Or by substituting $A_q$ and $A_0$ in the previous expression we can have the structure factor $S_q(t)$ in terms of $a$ 
\begin{equation}
\label{Sqintime}
 S_q(t) = \frac{2\tilde T\Delta t}{K_q(2+(1-2a)K_q)}\times F_q(t)
 \end{equation}
 with 
  \begin{equation}
 \label{functionoft}
  F_q(t)=1 - \left(\frac{1-aK_q}{1+(1-a)K_q}\right)^{2t/\Delta t}
 \end{equation}
 where $K_q$ encodes the time step $\Delta t$ and the Laplacian either with the exact expression $K_q= c\Delta t(2(1- \cos q))^{\alpha}$ or the approximation 
for small values of $q$: $K_q= c \Delta t q^{2\alpha}$ with $\alpha=\tfrac{1}{2}+\zeta$ for any roughness exponent 
$\zeta$.

 \begin{figure}[h!]
\centering  
\includegraphics{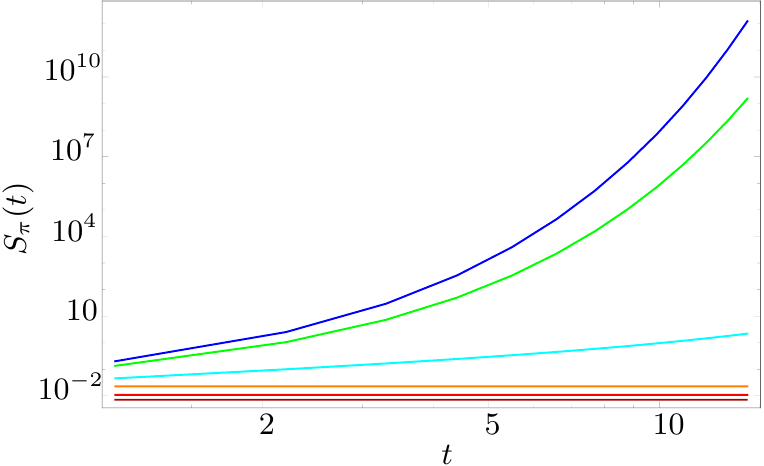}
\caption{\small  Stability of $S_q(t)$ as a function of time with $q=q^*=\pi$, $\Delta t=1.1>\Delta t_c$  and 
parameters $c=1$, $\zeta=1/2$. $S_{\pi}(t)$ is stable for all times for $a\leq1/2$ ($a=0,0.25,0.5$ in dark red, red and orange, respectively, from bottom up), 
however if $a$ becomes larger ($a=1,0.95, 0.75$ in blue, green and cyan, respectively, from top down) $S_{\pi}(t)$  loses stability}
\label{stabilityofSq}
\end{figure}

 The convergence of Eq.~\eqref{Sqintime} in the limit  $t\to\infty$ is guaranteed since $F_q(t)$ given by~\eqref{functionoft} converges to $1$ 
for any value of $a$ and $q$. In this limit the structure factor $S_q=S_q(t\to\infty)$ takes the 
following form
\begin{equation}
\label{Sqinfty}
S_q = \frac{\tilde T\Delta t}{K_q(1+\tfrac{(1-2a)}{2}K_q)}
\end{equation} 
with some particular cases corresponding to `anti-It\=o', Stratonovich and It\=o discretizations for  $a=0,1/2$ and $1$, 
respectively, which yield
\begin{align}
 \label{Sqgenerala0}
 S_q &=\frac{\tilde T \Delta t}{K_q(1+\frac{1}{2}K_q)} \;\;\;\;\;\;\;\;\;\;\;\;\;\;\;\;\;\;\;\;\;\;\;\;\;\;\;\;\;\;\; a=0\\
 \label{Sqgenerala12}
 S_q &=\frac{\tilde T \Delta t}{K_q}\;\;\;\;\;\;\;\;\;\;\;\;\;\;\;\;\;\;\;\;\;\;\;\;\;\;\;\;\;\;\;\;\;\;\;\;\;\;\;\;\;\;\;\; a=1/2\\
 \label{Sqgenerala1}
 S_q &=\frac{\tilde T \Delta t}{K_q(1-\frac{1}{2}K_q)} \;\;\;\;\;\;\;\;\;\;\;\;\;\;\;\;\;\;\;\;\;\;\;\;\;\;\;\;\;\;\; a=1
 \end{align} 
 where $K_q= c\Delta t(2(1- \cos q))^{\alpha}$ or  $K_q= c \Delta t q^{2\alpha}$ with $\alpha=\tfrac{1}{2}+\zeta$ for any roughness exponent $\zeta$.  In 
particular, the choice $a=1/2$ given by expression~\eqref{Sqgenerala12} cancels out the correction for $\Delta t$ in the 
structure factor. This constitutes one of our main results: the Stratonovitch discretization ($a=1/2$) is the discretization which minimizes the influence of 
the time step $\Delta t$ on the steady state. 
 
 Notice that in Eq.~\eqref{Sqinfty}, a wrong choice of $\Delta t$ can make the denominator equal to zero, this happens whenever $1+\tfrac{(1-2a)}{2}K_q=0$, 
i.e.~if 
\begin{equation}
\Delta t_c=\frac{2}{c\,(2a-1)\,(2(1- \cos q))^{\alpha}}
\label{eq:deltatc}
\end{equation}
for $a>1/2$ and $K_q=c\Delta t (2(1-\cos q))^{\alpha}$. For the explicit solution associated to the parameter $a=1$ (It\=o discretization) it is 
known from general experience that the solution of the EW equation is unstable if the time step is too large: this is indeed the Von Neumann stability 
criterion~\cite{press_numerical_2007}.
Our analysis hence provides a detailed understanding of this stability for any $a$:
the relation~\eqref{eq:deltatc} gives the critical value of $\Delta t$ for which $S_q$ becomes apparently negative as a result of the divergence of 
the geometric series~\eqref{eq:Sqgeomseries}. The expression~\eqref{eq:deltatc} thus represents the threshold above which the numerical procedure becomes 
unstable.

The smallest value that $\Delta t_c$ can take corresponds to the mode associated to $q^*=\pi$ for which
\begin{equation}
\label{deltatcritic}
 \Delta t_c(q^*)=\frac{2}{(2a-1)4^{\alpha}c}
\end{equation}
for $a>1/2$ and $\alpha=\tfrac{1}{2}+\zeta$ for any roughness exponent $\zeta$. 
This implies that the modes related to shorter distances are the first to become unstable. Fig.~\ref{stabilityofSq} shows the stability of $S_q(t)$ for 
different values of $a$ and a choice of $\Delta t=2.1>\Delta t_c$ with $q*=\pi$ and $c=1$, $\zeta=1/2$. The most important consequence of our analysis is that 
the structure factor $S_{\pi}(t)$ is stable at all times \emph{for any value $a\leq1/2$ of the discretization parameter}, while for larger values of $a$ it can 
lose stability for large enough time step $\Delta t$ given by expression~\eqref{deltatcritic}.

\subsection{\label{subsec:structureTrajectories} Structure factor from trajectorial probabilities}
By computing the stationary distribution  $P^{\text{st}}[u_q(t)]$ and comparing it with the probability of the reversed process we can obtain the 
structure factor $S_q$ in the stationary state. The dynamics of the process is not reversible since the normal process $u_q(t)$ and the same process reversed 
in time are described with It\=o ($a=1$) and anti-It\=o ($a=0$) discretization, respectively. We compare the probability of the process with the distribution 
of trajectories $P^{\text{traj}}[u_q(t),a]$ and the probability of the trajectories reversed in time $P^{\text{traj,R}}[u_q(t),1-a]$, whose temporal symmetry 
is conserved by the transformation $a\mapsto 1-a$. The probabilities of the process and of the trajectories compare as follows 
\begin{equation}
\label{stproba}
 \frac{P^{\text{st}}[u_q(t_f)]}{P^{\text{st}}[u_q(t_0)]}=\frac{P^{\text{traj}}[u_q(t),a]}{P^{\text{traj,R}}[u_q(t),1-a]}.
\end{equation}

The probability of a trajectory of $u$ will be deduced from that of the noise as
\begin{equation}
 \label{probaTraj}
 P^{\text{traj}}[u_q(t)]\propto \exp \left( -\frac{1}{2}\frac{1}{2\tilde T\Delta t} \sum_{t=0}^{t_{f}} \sum_q \eta_q(t)\eta_{-q}(t)  \right).
\end{equation}

In this case we focus on the implicit form of the equation \eqref{fourieruGen} with $a=0$, which it is widely used when working with numerical simulations due 
to its stability. This equation takes the following form
\begin{equation}
 \label{impliciteq}
 u_q(t)=\frac{u_q(t+\Delta t)+\sqrt{\tilde T \Delta t}\eta_q(t)}{1+\Delta t c q^2},
\end{equation}
where we use the approximation of the Laplacian for small $q$.

From Eqs.~\eqref{probaTraj} and \eqref{impliciteq} we can express $\eta_{q}(t)$ and $\eta_q^{\text{R}}(t)$ in terms of $u_q(t+\Delta t)$ and $u_q(t)$ to 
compute the right hand side of Eq.~\eqref{stproba}. For the forward trajectory we use $-\eta_q(t)=[u_q(t+\Delta t)-u_q(t)+\Delta t c q^2u_q(t)]/\sqrt{\tilde T 
\Delta t}$ and for the trajectory reversed in time we look at $-\eta_q^{\text{R}}(t)=[u_q(t+\Delta t)-u_q(t)-\Delta t c q^2u_q(t)]/\sqrt{\tilde T \Delta t}$ 
corresponding to $a=1$ in Eq.~\eqref{fourieruGen}. The factor in the sum on the right hand side in Eq.~\eqref{stproba} is found 
to be
\begin{align}
\nonumber
 [ \eta_q(t)\eta_{-q}(t)& - \eta_q^{\text{R}}(t)\eta_{-q}^{\text{R}}(t) ]=2\Delta t c \, q^2(1  + \tfrac{1}{2} \Delta t c q^2)\\
 \nonumber
\times & u_{q}(t+\Delta t)u_{-q}(t+\Delta t)-u_{q}(t)u_{-q}(t)\\
 \nonumber
 &\;\;\;\;\;\;\;\;\;\;\;\;\;\;\;\;\;\;\;\;\;\;\;=2\Delta t c \, \langle u_q^*(t)u_q(t) \rangle\\
 \label{diffetaq}
&\;\;\;\;\;\;\;\;\;\;\;\;\;\;\;\;\;\;\;\;\;\;\;\;\; \times[u_{q}(t')u_{-q}(t')]^{t+\Delta t}_t
\end{align}
where we denote $[u_{q}(t')u_{-q}(t')]^{t+\Delta t}_t=u_{q}(t+\Delta t)u_{-q}(t+\Delta t)-u_{q}(t)u_{-q}(t)$.  From the last expression~\eqref{diffetaq} we can 
identify the 
structure factor $S_q=S_q(t\to\infty)=\langle u_q^*(t)u_q(t) \rangle$. Therefore Eq.~\eqref{stproba} takes the form 
\begin{equation*}
 \frac{P^{\text{st}}[u_q(t_f)]}{P^{\text{st}}[u_q(t_0)]}=\exp\left( -\frac{1}{2} \sum_{t=0}^{t_f-\Delta t} \sum_q  S_q^{-1} [u_{q}(t')u_{-q}(t')]^{t+\Delta 
t}_t 
\right)
\end{equation*}
with the structure factor in the stationary state given by
\begin{equation}
 \label{sqcorrected}
 S_q=\frac{\tilde T}{cq^2(1+\tfrac{1}{2}c\Delta tq^2)}.
\end{equation}
Note that this expression for the structure factor differs from the common expression $S_q=1/q^2$ since it contains a 
correction term proportional to the time 
step $\Delta t$ induced by the time discretization. Equation~\eqref{sqcorrected} is in agreement with expression 
\eqref{Sqgenerala0} 
found in the previous section with the choice $a=0$ and $K_q=c\Delta t q^2$. 
 
 \subsection{\label{subsec:structureNumerics}Structure factor. Numerical simulations}
\begin{figure}[h!]
 \centering
\includegraphics{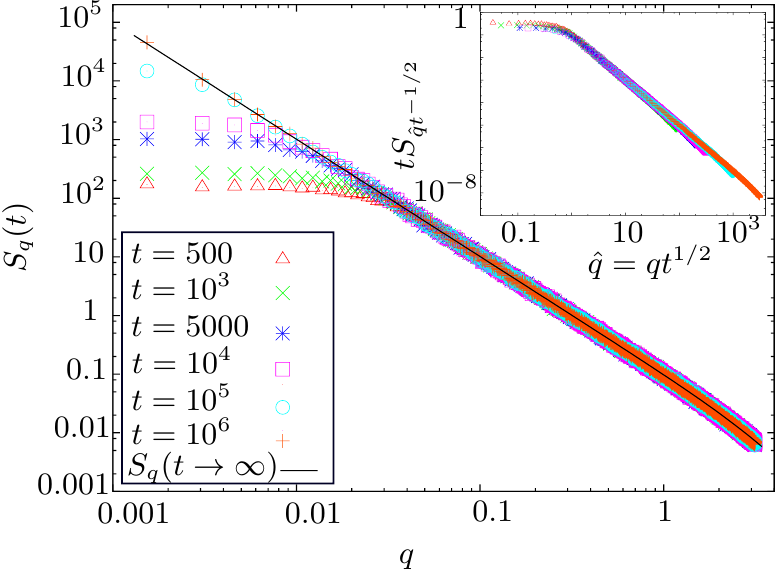}
\caption{Evolution of structure factor in time with  discretization parameter $a=0$ in the evolution equation \eqref{fourieruGen} with roughness exponent 
$\zeta=1/2$. The bending tail of $S_q(t)$ for $q$ close to $\pi$ is completely understood by means of the correction term proportional 
to $\Delta t$ and the approximate Laplacian encoded in the variable 
$K_q=c\Delta t q^2$ as derived in Eq.~\eqref{Sqgenerala0} (solid line). A dynamic scaling law is found to be $S_q=t^{-1} \hat{S}_{qt^{1/2}}$ (inset).}
\label{sqfortimesnumerics}
 \end{figure}

 For the evolution  of the structure factor $S_q(t)$ (see Fig.~\ref{sqfortimesnumerics}) we find a characteristic $q_c$, such that for $q<q_c$
 $S_q(t)$ saturates to a plateau whose value depends on $t$, while the large $q$ behavior is independent of $t$. This can also be compared with the analytic 
expression~\eqref{Sqintime}.
 
If we scale the structure factor as $S_{q}=t^{-1} \hat{S}_{qt^{1/2}}$ all the curves collapse in a single curve (see 
Fig.~\ref{sqfortimesnumerics}). This is in perfect agreement with the analytic expression found in Eq.~\eqref{Sqinfty}. 
It also shows that the non-steady relaxation of the interface is governed by a dynamical length growing 
as $L_{\text{dyn}}(t)\sim t^{1/2}$. In other words, we can write $S_{q} \equiv q^{2} \tilde{S}_{qL_{\text{dyn}}(t)}$, 
such that large lengthscales $q < 2 \pi/L_{\text{dyn}}(t)$ are out of equilibrium and retain 
memory of the flat initial condition $S_q \sim (2\pi/L_{\text{dyn}}(t))^{-2}$, while small lengthscales 
$q > 2 \pi/L_{\text{dyn}}(t)$ are equilibrated and display the characteristic equilibrium roughness exponent $S_q \sim q^{-2}$. Equilibration is thus expected 
for times $ t \gtrsim t_{\text{sat}} \sim L^2$, as $L_{\text{dyn}}(t) \to L$.

The tail of the structure factor $S_q(t)$ for $q$ close to $\pi$ (See Fig.~\ref{sqfortimesnumerics}) is controlled by the correction term proportional 
to $\Delta t$, and is strongly influenced by the choice of the parameter $a$ in the evolution equation \eqref{fourieruGen} and of the Laplacian 
encoded in the variable $K_q$ in Eq.~\eqref{Sqinfty}. In the simulations we used $a=0$ and $\tilde T=0.1$ and the approximate Laplacian $K_q=c\Delta t q^2$ 
with $c=1$ and $\Delta t=0.1$.

 \subsection{\label{subsec:density}Density of zeros}
\begin{figure}[h!]
\centering
\includegraphics{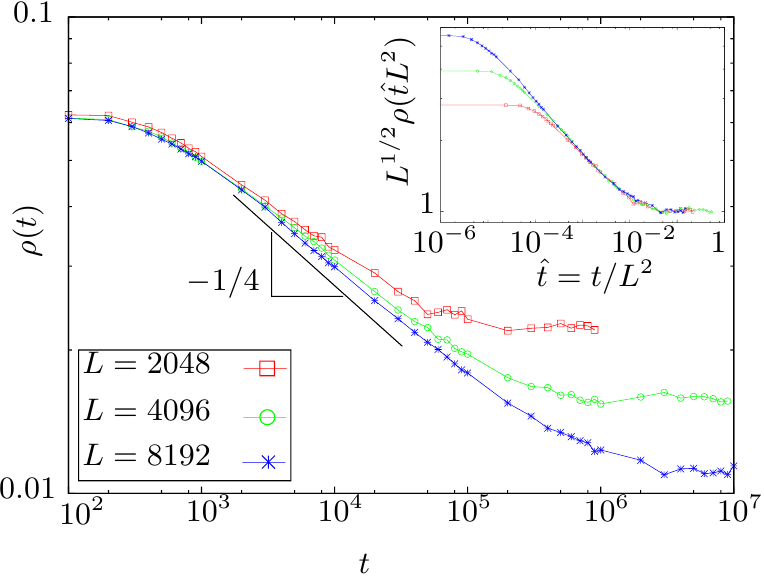}
\caption{\small  Evolution of density of zeros in time for different system sizes: $L=2048, 4096, 8192$ (in red, green and blue, 
respectively), averaged over 1000 realizations.  There is a characteristic time $t_c\sim L^2$ from which the density reaches a steady state. The scaling law 
for the density of zeros is found to be $\rho(L,t)=L^{-1/2}\hat{\rho}(t/L^{2})$.}
\label{densityvstime}
\end{figure}

Another observable is the non-stationary density $\rho$ of crossing zeros. 
The initial condition in the out-of-equilibrium state is a flat interface. Immediately 
after we let the interface  evolve, we observe that a large number of zeros appear. When the interface realizes it is finite, i.e.~at a saturation 
time $t_{\text{sat}}\sim L^2$, the density of zeros reaches a stationary state as shown in Fig.~\ref{densityvstime}. A scaling law can be found, it scales 
as $\rho(L,t)=L^{-1/2}\hat{\rho}(t/L^{2})$ for which a perfect collapse is found as shown in Fig.~\ref{densityvstime}. 

 We can extract a power law from the behavior of the density of zeros before the saturation time $t_{\text{sat}}$ which turns out to be $\rho(t)\sim 
t^{-1/4}$. 
This exponent is validated from the scaling found before since $\rho(L,t)=L^{-1/2}(t/L^{2})^{-1/4}=t^{-1/4}$. We can also see that the regime after the 
saturation time behaves as $\rho(L)\sim L^{-1/2}$. 
This is consistent with the finite-size scaling for $p(\ell)$ shown in Fig.~\ref{pdfell}. Since 
$\langle \ell \rangle \equiv \sum_{\ell=1}^{L} p(\ell) \ell \approx \int_0^L \ell^{-3/2} \ell d\ell \sim L^{1/2}$, 
we get that $\rho(L) = \langle \ell \rangle^{-1} \sim L^{-1/2}$. The density of zeros thus vanishes 
in the thermodynamic limit due to the infrared divergence of $\langle \ell \rangle$. This is 
directly related to the power-law decay exponent of $p(\ell)$ here $\gamma=3/2$, as predicted by the Sparre-Andersen theorem.
 
As discussed in relation with $S_q(t)$, the scaling law for $\rho(L,t)$ is consistent with a 
relaxation dominated by a single dynamical length $L_{\text{dyn}}(t) \sim t^{1/2}$. We can thus 
write $\rho(L,t)=L^{-1/2}\tilde{\rho}(L_{\text{dyn}}(t)/L)$, such that $\tilde{\rho}(x)\sim x^{1/2}$ for 
$x \ll 1$, 
corresponding to the non-steady regime, and $\tilde{\rho}(x)\sim \text{const.}$ 
for large $x \gg 1$, corresponding to the equilibrated regime. 
Comparing $\rho(t) \sim L_{\text{dyn}}(t)^{-1/2} \sim t^{-1/4}$, valid 
for intermediate times, with $\rho(t \to \infty) \sim L^{-1/2}$ we can see that 
finite-size scaling in the steady-state directly 
translates into the non-stationary finite-time scaling, 
by replacing $L \to L_{\text{dyn}}(t)$. 
In particular 
the power-law decay exponent $1/4$ in the density of zeros 
is related to the dynamical exponent $z=2$ in $L_{\text{dyn}}(t)\sim t^{1/z}$ 
and the Sparre-Andersen exponent $\gamma=3/2$ as $(2-\gamma)/z=1/4$. 
Interestingly, this let us predict that 
the non-stationary distribution of intervals can 
be expressed as $p(\ell,t) \sim \ell^{-3/2} \tilde{p}(\ell t^{-1/2})$ 
for $t<t_\text{sat}$.

\section{\label{sec:conclusions}Conclusions}
In this paper we have focused on fluctuating interfaces that belong to the EW universality class, i.e.~interfaces with roughness coefficient $\zeta=1/2$. We 
investigated the spatial first-passage probability which was obtained from the length of the intervals generated by the crossing zeros of the interface with 
respect to its average. The linearity of the Langevin equation ensures that the symmetry $u_x(t)\to-u_x(t)$ is conserved, therefore the persistence exponent 
for the positive and negative intervals is exactly the same \cite{constantin_spatial_2004}, i.e.~$\theta_{s}^{+}=\theta_{s}^{-}$. This justifies the fact that 
for stationary interfaces the distribution of intervals obtained numerically was obtained without making a distinction between the positive and the negative 
intervals. Regarding the distribution of intervals, both positive and negative, we have shown the agreement of the Sparre-Andersen theorem for random walks 
which measures the persistence in time with the spatial distribution of intervals for which the height in an interval is positively or negatively persistent. 
We also found a scaling function for such distribution related with the finite system size $L$ for which intervals larger than a 
certain $\ell_{\text{max}}\sim L/2$ are rare. We investigated the correlations between the intervals from which we obtained a scaling function with the 
system size $L$. The influence of periodic boundary conditions was also studied by means of the increments correlation of the interface itself.

Concerning the non-stationary regime we have presented a general discretization for the linear Langevin equation that simulates the evolution of fluctuating 
interfaces with roughness coefficient $\zeta$. An exact expression for the evolution of the structure factor was obtained and, surprisingly, it depends on  
the time step $\Delta t$ and on the discretization parameter $a$ even in the infinite-time limit. The correction in $\Delta t$ that it implies for the natural 
expression of the structure factor disappears for Stratonovich discretization corresponding to the choice of 
our parameter $a=1/2$. We have also found a relation that establishes the critical value of the time step $\Delta t_c$ needed as a function of the parameter 
$a$ so that stability of the simulation is guaranteed. Finally we study numerically the evolution of the structure factor $S_q(t)$ and the density $\rho$ of 
zeros . Regarding $S_q(t)$ two regimes were found, before a critical value $q_c\sim t^{1/2}$ we observe a plateau for small $q$ and for larger values of 
$q>q_c$ $S_q(t)$ presents a power-law decay which goes as $\sim q^2$. The stationary limit found numerically is in perfect agreement with the analytic 
expression found for $S_q(t\to\infty)$. For the density of zeros $\rho$ two regimes were found as well. Before a saturation time that scales as 
$t_{\text{sat}}\sim L^2$ the density of zeros follows a power-law with a decay that goes as $\sim t^{-1/4}$, which is in perfect agreement with the scaling of 
reaction diffusion processes found in the literature \cite{toussaint_particleantiparticle_1983}, and for times larger than $t_{\text{sat}}$ the density reaches 
a stationary state as expected.

\section*{Acknowledgements}
The authors acknowledge the ECOS project A12E05 which allowed this collaboration. A.B.K wishes to thank the European Union EMMCSS visiting scientist program 
and acknowledges partial support from Projects PIP11220090100051 and PIP11220120100250CO (CONICET). V.L. acknowledges the Coopinter project EDC25533 for 
partial funding. A.L.Z wishes to thank the hospitality of the Condensed Matter Theory Group at Centro At\'omico Bariloche, S.C. de Bariloche, Argentina, and 
acknowledges funding from the Mexican National Council for Science and Technology (CONACyT-M\'exico) through a scholarship to pursue graduate studies at the
Laboratoire des Probabilit\'es et Mod\`eles Al\'eatoires (LPMA), Paris, France.

\bibliographystyle{apsrev}
\bibliography{Bibliography}

\begin{thebibliography}{39}
\expandafter\ifx\csname natexlab\endcsname\relax\def\natexlab#1{#1}\fi
\expandafter\ifx\csname bibnamefont\endcsname\relax
  \def\bibnamefont#1{#1}\fi
\expandafter\ifx\csname bibfnamefont\endcsname\relax
  \def\bibfnamefont#1{#1}\fi
\expandafter\ifx\csname citenamefont\endcsname\relax
  \def\citenamefont#1{#1}\fi
\expandafter\ifx\csname url\endcsname\relax
  \def\url#1{\texttt{#1}}\fi
\expandafter\ifx\csname urlprefix\endcsname\relax\def\urlprefix{URL }\fi
\providecommand{\bibinfo}[2]{#2}
\providecommand{\eprint}[2][]{\url{#2}}

\bibitem[{\citenamefont{Redner}(2001)}]{redner_guide_2001}
\bibinfo{author}{\bibfnamefont{S.}~\bibnamefont{Redner}},
  \emph{\bibinfo{title}{A {Guide} to {First}-{Passage} {Processes}}}
  (\bibinfo{publisher}{Cambridge University Press}, \bibinfo{year}{2001}), ISBN
  \bibinfo{isbn}{978-0-521-65248-3}.

\bibitem[{\citenamefont{Montroll and Weiss}(1965)}]{montroll_random_1965}
\bibinfo{author}{\bibfnamefont{E.~W.} \bibnamefont{Montroll}} \bibnamefont{and}
  \bibinfo{author}{\bibfnamefont{G.~H.} \bibnamefont{Weiss}},
  \bibinfo{journal}{Journal of Mathematical Physics}
  \textbf{\bibinfo{volume}{6}}, \bibinfo{pages}{167} (\bibinfo{year}{1965}),
  ISSN \bibinfo{issn}{0022-2488, 1089-7658},
  \urlprefix\url{http://scitation.aip.org/content/aip/journal/jmp/6/2/10.1063/1.1704269}.

\bibitem[{\citenamefont{McKean}(1962)}]{mckean_winding_1962}
\bibinfo{author}{\bibfnamefont{H.~P.~J.} \bibnamefont{McKean}},
  \bibinfo{journal}{J. Math. Kyoto Univ.} \textbf{\bibinfo{volume}{2}},
  \bibinfo{pages}{227} (\bibinfo{year}{1962}), ISSN \bibinfo{issn}{0023-608X},
  \urlprefix\url{http://projecteuclid.org/euclid.kjm/1250524936}.

\bibitem[{\citenamefont{Sinai}(1992{\natexlab{a}})}]{sinai_distribution_1992}
\bibinfo{author}{\bibfnamefont{Y.~G.} \bibnamefont{Sinai}},
  \bibinfo{journal}{Theor Math Phys} \textbf{\bibinfo{volume}{90}},
  \bibinfo{pages}{219} (\bibinfo{year}{1992}{\natexlab{a}}), ISSN
  \bibinfo{issn}{0040-5779, 1573-9333},
  \urlprefix\url{http://link.springer.com/article/10.1007/BF01036528}.

\bibitem[{\citenamefont{Sinai}(1992{\natexlab{b}})}]{sinai_statistics_1992}
\bibinfo{author}{\bibfnamefont{Y.~G.} \bibnamefont{Sinai}},
  \bibinfo{journal}{Commun.Math. Phys.} \textbf{\bibinfo{volume}{148}},
  \bibinfo{pages}{601} (\bibinfo{year}{1992}{\natexlab{b}}), ISSN
  \bibinfo{issn}{0010-3616, 1432-0916},
  \urlprefix\url{http://link.springer.com/article/10.1007/BF02096550}.

\bibitem[{\citenamefont{Derrida et~al.}(1994)\citenamefont{Derrida, Bray, and
  Godreche}}]{derrida_non-trivial_1994}
\bibinfo{author}{\bibfnamefont{B.}~\bibnamefont{Derrida}},
  \bibinfo{author}{\bibfnamefont{A.~J.} \bibnamefont{Bray}}, \bibnamefont{and}
  \bibinfo{author}{\bibfnamefont{C.}~\bibnamefont{Godreche}},
  \bibinfo{journal}{J. Phys. A: Math. Gen.} \textbf{\bibinfo{volume}{27}},
  \bibinfo{pages}{L357} (\bibinfo{year}{1994}), ISSN \bibinfo{issn}{0305-4470},
  \urlprefix\url{http://stacks.iop.org/0305-4470/27/i=11/a=002}.

\bibitem[{\citenamefont{Bray}(1994)}]{bray_theory_1994}
\bibinfo{author}{\bibfnamefont{A.~J.} \bibnamefont{Bray}},
  \bibinfo{journal}{Advances in Physics} \textbf{\bibinfo{volume}{43}},
  \bibinfo{pages}{357} (\bibinfo{year}{1994}), ISSN \bibinfo{issn}{0001-8732},
  \urlprefix\url{http://dx.doi.org/10.1080/00018739400101505}.

\bibitem[{\citenamefont{Hilhorst}(2000)}]{hilhorst_persistence_2000}
\bibinfo{author}{\bibfnamefont{H.~J.} \bibnamefont{Hilhorst}},
  \bibinfo{journal}{Physica A: Statistical Mechanics and its Applications}
  \textbf{\bibinfo{volume}{277}}, \bibinfo{pages}{124} (\bibinfo{year}{2000}),
  ISSN \bibinfo{issn}{0378-4371},
  \urlprefix\url{http://www.sciencedirect.com/science/article/pii/S0378437199005099}.

\bibitem[{\citenamefont{Dougherty et~al.}(2002)\citenamefont{Dougherty,
  Lyubinetsky, Williams, Constantin, Dasgupta, and
  Sarma}}]{dougherty_experimental_2002}
\bibinfo{author}{\bibfnamefont{D.~B.} \bibnamefont{Dougherty}},
  \bibinfo{author}{\bibfnamefont{I.}~\bibnamefont{Lyubinetsky}},
  \bibinfo{author}{\bibfnamefont{E.~D.} \bibnamefont{Williams}},
  \bibinfo{author}{\bibfnamefont{M.}~\bibnamefont{Constantin}},
  \bibinfo{author}{\bibfnamefont{C.}~\bibnamefont{Dasgupta}}, \bibnamefont{and}
  \bibinfo{author}{\bibfnamefont{S.}~\bibnamefont{Sarma}},
  \bibinfo{journal}{Phys. Rev. Lett.} \textbf{\bibinfo{volume}{89}},
  \bibinfo{pages}{136102} (\bibinfo{year}{2002}),
  \urlprefix\url{http://link.aps.org/doi/10.1103/PhysRevLett.89.136102}.

\bibitem[{\citenamefont{Krug}(1997)}]{krug_origins_1997}
\bibinfo{author}{\bibfnamefont{J.}~\bibnamefont{Krug}},
  \bibinfo{journal}{Advances in Physics} \textbf{\bibinfo{volume}{46}},
  \bibinfo{pages}{139} (\bibinfo{year}{1997}), ISSN \bibinfo{issn}{0001-8732},
  \urlprefix\url{http://dx.doi.org/10.1080/00018739700101498}.

\bibitem[{\citenamefont{Kallabis and Krug}(1999)}]{kallabis_persistence_1999}
\bibinfo{author}{\bibfnamefont{H.}~\bibnamefont{Kallabis}} \bibnamefont{and}
  \bibinfo{author}{\bibfnamefont{J.}~\bibnamefont{Krug}},
  \bibinfo{journal}{Europhysics Letters (EPL)} \textbf{\bibinfo{volume}{45}},
  \bibinfo{pages}{20} (\bibinfo{year}{1999}), ISSN \bibinfo{issn}{0295-5075,
  1286-4854},
  \urlprefix\url{http://stacks.iop.org/0295-5075/45/i=1/a=020?key=crossref.14d78e6d02dec0e02f71c54cfddbc3a9}.

\bibitem[{\citenamefont{Godr\`eche}(1991)}]{godreche_solids_1991}
\bibinfo{author}{\bibfnamefont{C.}~\bibnamefont{Godr\`eche}},
  \emph{\bibinfo{title}{Solids {Far} from {Equilibrium}}}
  (\bibinfo{publisher}{Cambridge University Press}, \bibinfo{year}{1991}), ISBN
  \bibinfo{isbn}{978-0-521-41170-7}.

\bibitem[{\citenamefont{Bray et~al.}(2013)\citenamefont{Bray, Majumdar, and
  Schehr}}]{bray_persistence_2013}
\bibinfo{author}{\bibfnamefont{A.~J.} \bibnamefont{Bray}},
  \bibinfo{author}{\bibfnamefont{S.~N.} \bibnamefont{Majumdar}},
  \bibnamefont{and} \bibinfo{author}{\bibfnamefont{G.}~\bibnamefont{Schehr}},
  \bibinfo{journal}{Advances in Physics} \textbf{\bibinfo{volume}{62}},
  \bibinfo{pages}{225} (\bibinfo{year}{2013}), ISSN \bibinfo{issn}{0001-8732},
  \urlprefix\url{http://dx.doi.org/10.1080/00018732.2013.803819}.

\bibitem[{\citenamefont{Majumdar and Bray}(2001)}]{majumdar_spatial_2001}
\bibinfo{author}{\bibfnamefont{S.~N.} \bibnamefont{Majumdar}} \bibnamefont{and}
  \bibinfo{author}{\bibfnamefont{A.~J.} \bibnamefont{Bray}},
  \bibinfo{journal}{Phys. Rev. Lett.} \textbf{\bibinfo{volume}{86}},
  \bibinfo{pages}{3700} (\bibinfo{year}{2001}),
  \urlprefix\url{http://link.aps.org/doi/10.1103/PhysRevLett.86.3700}.

\bibitem[{\citenamefont{Constantin
  et~al.}(2004{\natexlab{a}})\citenamefont{Constantin, Das~Sarma, and
  Dasgupta}}]{constantin_spatial_2004}
\bibinfo{author}{\bibfnamefont{M.}~\bibnamefont{Constantin}},
  \bibinfo{author}{\bibfnamefont{S.}~\bibnamefont{Das~Sarma}},
  \bibnamefont{and} \bibinfo{author}{\bibfnamefont{C.}~\bibnamefont{Dasgupta}},
  \bibinfo{journal}{Phys. Rev. E} \textbf{\bibinfo{volume}{69}},
  \bibinfo{pages}{051603} (\bibinfo{year}{2004}{\natexlab{a}}),
  \urlprefix\url{http://link.aps.org/doi/10.1103/PhysRevE.69.051603}.

\bibitem[{\citenamefont{Majumdar and Dasgupta}(2006)}]{majumdar_spatial_2006}
\bibinfo{author}{\bibfnamefont{S.~N.} \bibnamefont{Majumdar}} \bibnamefont{and}
  \bibinfo{author}{\bibfnamefont{C.}~\bibnamefont{Dasgupta}},
  \bibinfo{journal}{Phys. Rev. E} \textbf{\bibinfo{volume}{73}},
  \bibinfo{pages}{011602} (\bibinfo{year}{2006}),
  \urlprefix\url{http://link.aps.org/doi/10.1103/PhysRevE.73.011602}.

\bibitem[{\citenamefont{Sire}(2008)}]{sire_crossing_2008}
\bibinfo{author}{\bibfnamefont{C.}~\bibnamefont{Sire}}, \bibinfo{journal}{Phys.
  Rev. E} \textbf{\bibinfo{volume}{78}}, \bibinfo{pages}{011121}
  (\bibinfo{year}{2008}),
  \urlprefix\url{http://link.aps.org/doi/10.1103/PhysRevE.78.011121}.

\bibitem[{\citenamefont{Alemany and ben
  Avraham}(1995)}]{alemany_inter-particle_1995}
\bibinfo{author}{\bibfnamefont{P.~A.} \bibnamefont{Alemany}} \bibnamefont{and}
  \bibinfo{author}{\bibfnamefont{D.}~\bibnamefont{ben Avraham}},
  \bibinfo{journal}{Physics Letters A} \textbf{\bibinfo{volume}{206}},
  \bibinfo{pages}{18} (\bibinfo{year}{1995}), ISSN \bibinfo{issn}{0375-9601},
  \urlprefix\url{http://www.sciencedirect.com/science/article/pii/037596019500625D}.

\bibitem[{\citenamefont{Derrida et~al.}(1996)\citenamefont{Derrida, Hakim, and
  Zeitak}}]{derrida_persistent_1996}
\bibinfo{author}{\bibfnamefont{B.}~\bibnamefont{Derrida}},
  \bibinfo{author}{\bibfnamefont{V.}~\bibnamefont{Hakim}}, \bibnamefont{and}
  \bibinfo{author}{\bibfnamefont{R.}~\bibnamefont{Zeitak}},
  \bibinfo{journal}{Phys. Rev. Lett.} \textbf{\bibinfo{volume}{77}},
  \bibinfo{pages}{2871} (\bibinfo{year}{1996}),
  \urlprefix\url{http://link.aps.org/doi/10.1103/PhysRevLett.77.2871}.

\bibitem[{\citenamefont{Majumdar et~al.}(1996)\citenamefont{Majumdar, Sire,
  Bray, and Cornell}}]{majumdar_nontrivial_1996}
\bibinfo{author}{\bibfnamefont{S.~N.} \bibnamefont{Majumdar}},
  \bibinfo{author}{\bibfnamefont{C.}~\bibnamefont{Sire}},
  \bibinfo{author}{\bibfnamefont{A.~J.} \bibnamefont{Bray}}, \bibnamefont{and}
  \bibinfo{author}{\bibfnamefont{S.~J.} \bibnamefont{Cornell}},
  \bibinfo{journal}{Phys. Rev. Lett.} \textbf{\bibinfo{volume}{77}},
  \bibinfo{pages}{2867} (\bibinfo{year}{1996}),
  \urlprefix\url{http://link.aps.org/doi/10.1103/PhysRevLett.77.2867}.

\bibitem[{\citenamefont{Constantin
  et~al.}(2004{\natexlab{b}})\citenamefont{Constantin, Dasgupta, Chatraphorn,
  Majumdar, and Das~Sarma}}]{constantin_persistence_2004}
\bibinfo{author}{\bibfnamefont{M.}~\bibnamefont{Constantin}},
  \bibinfo{author}{\bibfnamefont{C.}~\bibnamefont{Dasgupta}},
  \bibinfo{author}{\bibfnamefont{P.~P.} \bibnamefont{Chatraphorn}},
  \bibinfo{author}{\bibfnamefont{S.~N.} \bibnamefont{Majumdar}},
  \bibnamefont{and}
  \bibinfo{author}{\bibfnamefont{S.}~\bibnamefont{Das~Sarma}},
  \bibinfo{journal}{Phys. Rev. E} \textbf{\bibinfo{volume}{69}},
  \bibinfo{pages}{061608} (\bibinfo{year}{2004}{\natexlab{b}}),
  \urlprefix\url{http://link.aps.org/doi/10.1103/PhysRevE.69.061608}.

\bibitem[{\citenamefont{McFadden}(1956)}]{mcfadden_axis-crossing_1956}
\bibinfo{author}{\bibfnamefont{J.}~\bibnamefont{McFadden}},
  \bibinfo{journal}{IRE Transactions on Information Theory}
  \textbf{\bibinfo{volume}{2}}, \bibinfo{pages}{146} (\bibinfo{year}{1956}),
  ISSN \bibinfo{issn}{0096-1000}.

\bibitem[{\citenamefont{McFadden}(1958)}]{mcfadden_axis-crossing_1958}
\bibinfo{author}{\bibfnamefont{J.}~\bibnamefont{McFadden}},
  \bibinfo{journal}{IRE Transactions on Information Theory}
  \textbf{\bibinfo{volume}{4}}, \bibinfo{pages}{14} (\bibinfo{year}{1958}),
  ISSN \bibinfo{issn}{0096-1000}.

\bibitem[{\citenamefont{Lamperti}(1962)}]{lamperti_semi-stable_1962}
\bibinfo{author}{\bibfnamefont{J.}~\bibnamefont{Lamperti}},
  \bibinfo{journal}{Transactions of the American Mathematical Society}
  \textbf{\bibinfo{volume}{104}}, \bibinfo{pages}{62} (\bibinfo{year}{1962}),
  ISSN \bibinfo{issn}{0002-9947},
  \urlprefix\url{http://www.jstor.org/stable/1993933}.

\bibitem[{\citenamefont{Majumdar and Sire}(1996)}]{majumdar_survival_1996}
\bibinfo{author}{\bibfnamefont{S.~N.} \bibnamefont{Majumdar}} \bibnamefont{and}
  \bibinfo{author}{\bibfnamefont{C.}~\bibnamefont{Sire}},
  \bibinfo{journal}{Phys. Rev. Lett.} \textbf{\bibinfo{volume}{77}},
  \bibinfo{pages}{1420} (\bibinfo{year}{1996}),
  \urlprefix\url{http://link.aps.org/doi/10.1103/PhysRevLett.77.1420}.

\bibitem[{\citenamefont{Oerding et~al.}(1997)\citenamefont{Oerding, Cornell,
  and Bray}}]{oerding_non-markovian_1997}
\bibinfo{author}{\bibfnamefont{K.}~\bibnamefont{Oerding}},
  \bibinfo{author}{\bibfnamefont{S.~J.} \bibnamefont{Cornell}},
  \bibnamefont{and} \bibinfo{author}{\bibfnamefont{A.~J.} \bibnamefont{Bray}},
  \bibinfo{journal}{Phys. Rev. E} \textbf{\bibinfo{volume}{56}},
  \bibinfo{pages}{R25} (\bibinfo{year}{1997}),
  \urlprefix\url{http://link.aps.org/doi/10.1103/PhysRevE.56.R25}.

\bibitem[{\citenamefont{Krug et~al.}(1997)\citenamefont{Krug, Kallabis,
  Majumdar, Cornell, Bray, and Sire}}]{krug_persistence_1997}
\bibinfo{author}{\bibfnamefont{J.}~\bibnamefont{Krug}},
  \bibinfo{author}{\bibfnamefont{H.}~\bibnamefont{Kallabis}},
  \bibinfo{author}{\bibfnamefont{S.~N.} \bibnamefont{Majumdar}},
  \bibinfo{author}{\bibfnamefont{S.~J.} \bibnamefont{Cornell}},
  \bibinfo{author}{\bibfnamefont{A.~J.} \bibnamefont{Bray}}, \bibnamefont{and}
  \bibinfo{author}{\bibfnamefont{C.}~\bibnamefont{Sire}},
  \bibinfo{journal}{Phys. Rev. E} \textbf{\bibinfo{volume}{56}},
  \bibinfo{pages}{2702} (\bibinfo{year}{1997}),
  \urlprefix\url{http://link.aps.org/doi/10.1103/PhysRevE.56.2702}.

\bibitem[{\citenamefont{Zamorategui et~al.}(2016)\citenamefont{Zamorategui,
  Lecomte, and Kolton}}]{futureworkonnonmarkovian}
\bibinfo{author}{\bibfnamefont{A.~L.} \bibnamefont{Zamorategui}},
  \bibinfo{author}{\bibfnamefont{V.}~\bibnamefont{Lecomte}}, \bibnamefont{and}
  \bibinfo{author}{\bibfnamefont{A.~B.} \bibnamefont{Kolton}},
  \bibinfo{journal}{In preparation}  (\bibinfo{year}{2016}).

\bibitem[{\citenamefont{Chakraborty and
  Bhattacharjee}(2007)}]{chakraborty_finite-size_2007}
\bibinfo{author}{\bibfnamefont{D.}~\bibnamefont{Chakraborty}} \bibnamefont{and}
  \bibinfo{author}{\bibfnamefont{J.~K.} \bibnamefont{Bhattacharjee}},
  \bibinfo{journal}{Phys. Rev. E} \textbf{\bibinfo{volume}{75}},
  \bibinfo{pages}{011111} (\bibinfo{year}{2007}),
  \urlprefix\url{http://link.aps.org/doi/10.1103/PhysRevE.75.011111}.

\bibitem[{\citenamefont{Majumdar and Comtet}(2005)}]{majumdar_airy_2005}
\bibinfo{author}{\bibfnamefont{S.~N.} \bibnamefont{Majumdar}} \bibnamefont{and}
  \bibinfo{author}{\bibfnamefont{A.}~\bibnamefont{Comtet}}, \bibinfo{journal}{J
  Stat Phys} \textbf{\bibinfo{volume}{119}}, \bibinfo{pages}{777}
  (\bibinfo{year}{2005}), ISSN \bibinfo{issn}{0022-4715, 1572-9613},
  \urlprefix\url{http://link.springer.com/article/10.1007/s10955-005-3022-4}.

\bibitem[{\citenamefont{Schehr and Majumdar}(2006)}]{schehr_universal_2006}
\bibinfo{author}{\bibfnamefont{G.}~\bibnamefont{Schehr}} \bibnamefont{and}
  \bibinfo{author}{\bibfnamefont{S.~N.} \bibnamefont{Majumdar}},
  \bibinfo{journal}{Phys. Rev. E} \textbf{\bibinfo{volume}{73}},
  \bibinfo{pages}{056103} (\bibinfo{year}{2006}),
  \urlprefix\url{http://link.aps.org/doi/10.1103/PhysRevE.73.056103}.

\bibitem[{\citenamefont{Andersen}(1953)}]{andersen_fluctuations_1953}
\bibinfo{author}{\bibfnamefont{E.~S.} \bibnamefont{Andersen}},
  \bibinfo{journal}{Math. Scand} \textbf{\bibinfo{volume}{1}},
  \bibinfo{pages}{1954} (\bibinfo{year}{1953}).

\bibitem[{\citenamefont{Press}(2007)}]{press_numerical_2007}
\bibinfo{author}{\bibfnamefont{W.~H.} \bibnamefont{Press}},
  \emph{\bibinfo{title}{Numerical {Recipes} 3rd {Edition}: {The} {Art} of
  {Scientific} {Computing}}} (\bibinfo{publisher}{Cambridge University Press},
  \bibinfo{year}{2007}), ISBN \bibinfo{isbn}{978-0-521-88068-8}.

\bibitem[{\citenamefont{Ito}(1944)}]{ito_109._1944}
\bibinfo{author}{\bibfnamefont{K.}~\bibnamefont{Ito}},
  \bibinfo{journal}{Proceedings of the Imperial Academy}
  \textbf{\bibinfo{volume}{20}}, \bibinfo{pages}{519} (\bibinfo{year}{1944}).

\bibitem[{\citenamefont{Stratonovich}(1967)}]{stratonovich_topics_1967}
\bibinfo{author}{\bibfnamefont{R.~L.} \bibnamefont{Stratonovich}},
  \emph{\bibinfo{title}{Topics {In} the {Theory} of {Random} {Noise}}}
  (\bibinfo{publisher}{CRC Press}, \bibinfo{year}{1967}), ISBN
  \bibinfo{isbn}{978-0-677-00790-8}.

\bibitem[{\citenamefont{Frachebourg et~al.}(1995)\citenamefont{Frachebourg,
  Ispolatov, and Krapivsky}}]{frachebourg_extremal_1995}
\bibinfo{author}{\bibfnamefont{L.}~\bibnamefont{Frachebourg}},
  \bibinfo{author}{\bibfnamefont{I.}~\bibnamefont{Ispolatov}},
  \bibnamefont{and} \bibinfo{author}{\bibfnamefont{P.~L.}
  \bibnamefont{Krapivsky}}, \bibinfo{journal}{Phys. Rev. E}
  \textbf{\bibinfo{volume}{52}}, \bibinfo{pages}{R5727} (\bibinfo{year}{1995}),
  \urlprefix\url{http://link.aps.org/doi/10.1103/PhysRevE.52.R5727}.

\bibitem[{\citenamefont{Godr{\`e}che et~al.}(2015)\citenamefont{Godr{\`e}che,
  Majumdar, and Schehr}}]{godreche2015record}
\bibinfo{author}{\bibfnamefont{C.}~\bibnamefont{Godr{\`e}che}},
  \bibinfo{author}{\bibfnamefont{S.~N.} \bibnamefont{Majumdar}},
  \bibnamefont{and} \bibinfo{author}{\bibfnamefont{G.}~\bibnamefont{Schehr}},
  \bibinfo{journal}{Journal of Statistical Mechanics: Theory and Experiment}
  \textbf{\bibinfo{volume}{2015}}, \bibinfo{pages}{P07026}
  (\bibinfo{year}{2015}).

\bibitem[{\citenamefont{Krauth}(2006)}]{krauth_statistical_2006}
\bibinfo{author}{\bibfnamefont{W.}~\bibnamefont{Krauth}},
  \emph{\bibinfo{title}{Statistical {Mechanics}: {Algorithms} and
  {Computations}}} (\bibinfo{publisher}{Oxford University Press, UK},
  \bibinfo{year}{2006}), ISBN \bibinfo{isbn}{978-0-19-851535-7}.

\bibitem[{\citenamefont{Toussaint and
  Wilczek}(1983)}]{toussaint_particleantiparticle_1983}
\bibinfo{author}{\bibfnamefont{D.}~\bibnamefont{Toussaint}} \bibnamefont{and}
  \bibinfo{author}{\bibfnamefont{F.}~\bibnamefont{Wilczek}},
  \bibinfo{journal}{The Journal of Chemical Physics}
  \textbf{\bibinfo{volume}{78}}, \bibinfo{pages}{2642} (\bibinfo{year}{1983}),
  ISSN \bibinfo{issn}{0021-9606, 1089-7690},
  \urlprefix\url{http://scitation.aip.org/content/aip/journal/jcp/78/5/10.1063/1.445022}.

\end{thebibliography}

\end{document}